\documentclass[%
 preprint,
 amsmath,amssymb,
 aps,
 prd,
longbibliography
]{revtex4-2}

\usepackage{graphicx}
\usepackage{dcolumn}
\usepackage{bm}
\usepackage[bookmarks=true,colorlinks=true,linkcolor=blue,unicode=true]{hyperref}
\usepackage[mathlines]{lineno}
\usepackage{xcolor} 
\usepackage{color} 
\usepackage{slashed}

\newcounter{YJC}

\begin{document}

\title{Mesonic screening correlators in an external imaginary electric field at finite temperature}

\author{Ji-Chong Yang}
\email{yangjichong@lnnu.edu.cn}
\thanks{corresponding author}
\author{Zhan Zhao}
\email{zhaozhan061300@163.com}
\author{Xiang-Ning Li}
\email{2788249040@qq.com}
\author{Wen-Wen Li}
\email{lww1514148@163.com}

\affiliation{Department of Physics, Liaoning Normal University, Dalian 116029, China}
\affiliation{Center for Theoretical and Experimental High Energy Physics, Liaoning Normal University, Dalian 116029, China}

\date{\today}

\begin{abstract}
External electromagnetic fields provide a useful probe of QCD matter, but real electric fields are hindered by the sign problem, motivating studies with imaginary electric fields. 
We investigate mesonic screening correlators in lattice QCD at finite temperature in the presence of such a background using staggered fermions. 
At low temperature, scalar screening masses increase with the field strength, while pseudo-scalar masses remain largely unchanged, and charge-asymmetric channels show additional structure. 
At high temperature, the correlators exhibit clear spatial oscillations with frequencies set by the quark electric charges. 
These results demonstrate nontrivial modifications of screening properties induced by external electric fields.
\end{abstract}

\maketitle


\section{\label{sec:sec1}Introduction}

The response of strongly interacting matter to external electromagnetic fields has been extensively studied in recent years, motivated in part by the realization that intense fields are generated in non-central heavy-ion collisions~\cite{Kharzeev:2007jp,Skokov:2009qp,Deng:2012pc}. 
These fields can reach magnitudes comparable to typical QCD scales and provide a unique probe of the structure and dynamics of the quark-gluon plasma. 
In particular, electromagnetic backgrounds have been used to investigate anomalous transport phenomena such as the chiral magnetic effect~\cite{Kharzeev:2004ey,Fukushima:2008xe}, as well as more general aspects of QCD thermodynamics and symmetry properties~\cite{Andersen:2014xxa}.

The effects of magnetic fields on QCD have been widely explored using both effective models and lattice simulations. 
Lattice studies have established, for example, the phenomenon of inverse magnetic catalysis near the chiral crossover temperature~\cite{Bali:2011qj,Bali:2012zg}.
Lattice studies have examined the behavior of meson correlators and screening masses in magnetic backgrounds, including recent high-precision investigations at physical quark masses which reveal a nontrivial interplay between screening properties and chiral dynamics~\cite{Braguta:2011hq,Hidaka:2012mz,Ding:2025pbu}. 
The results provide a natural motivation to explore how external electric fields modify screening correlators, because in the relativistic heavy-ion collisions the electric field strength can be at the same order as the magnetic field~\cite{Bzdak:2011yy,Bloczynski:2012en,Deng:2012pc,Hirono:2012rt,Deng:2014uja,Voronyuk:2014rna}.
In contrast, the study of electric fields is significantly more challenging due to the sign problem. 
A real electric field renders the fermion determinant complex, preventing direct application of importance sampling methods, except for the case where the $u$ and $d$ quarks have opposite charges of equal magnitude~\cite{Yamamoto:2012bd}. 
A common approach to circumvent this difficulty is to introduce an imaginary electric field, for which the Euclidean action remains real and standard lattice techniques can be employed~\cite{Shintani:2006xr,Alexandru:2008sj,DElia:2012ifm,Christensen:2004ca,Engelhardt:2007ub,Endrodi:2021qxz}. 
Although analytic continuation is required to relate such results to real electric fields, this method provides a practical framework for exploring qualitative features of QCD in external electric backgrounds.

At finite temperature, spatial correlation functions (screening correlators) offer a powerful probe of in-medium physics~\cite{Detar:1987hib}. 
The associated screening masses characterize the spatial decay of correlations and encode information about the relevant degrees of freedom. 
In the confined phase, screening correlators are dominated by hadronic excitations, while in the deconfined phase they reflect the propagation of quark and gluon degrees of freedom, approaching the free-field expectation $m \sim 2\pi T$ at high temperature~\cite{Florkowski:1993bq,Laine:2003bd,Alberico:2007yj}. 
Lattice studies have extensively investigated screening masses and their relation to chiral symmetry restoration~\cite{Cheng:2010fe}.

External electromagnetic fields provide an additional handle to probe the structure of screening states. 
In particular, electric fields couple directly to quark charges and can induce nontrivial modifications of correlation functions. 
In this work, we investigate mesonic screening correlators in the presence of an external~(imaginary) electric field using lattice QCD. 
We focus on the behavior of various spin channels constructed from local staggered fermion bilinears. 
Special attention is paid to the emergence of oscillatory behavior in certain channels, which is absent in the zero-field case. 
We analyze the dependence of this behavior on the electric field strength and on the quark charges, and explore its implications for the underlying dynamics. 
The observed patterns suggest that the external field induces nontrivial phase structures in the correlators, which become particularly pronounced at high temperature where screening states are less bound.

This paper is organized as follows. 
In Sec.~\ref{sec:sec2} we describe the lattice setup and the implementation of the external electric field. 
Numerical results for chiral condensate, Polyakov loop, and screening correlators and effective masses are presented in Sec.~\ref{sec:sec3}, followed by a discussion of the observed oscillatory behavior and its possible interpretation. 
We conclude in Sec.~\ref{sec:sec4} with a summary and outlook.

\section{\label{sec:sec2}The model with external electric field}

We consider an external electric field oriented along the $x$-direction, $\mathbf{E} = (E_x,0,0)$. 
In the axial gauge $A^{\text{EM}}_x = 0$, the associated gauge field can be expressed as $A^{\text{EM}}_\mu = (-E_x x, 0, 0, 0)$, such that $\mathbf{E} = (F^{\text{EM}}_{tx}, F^{\text{EM}}_{ty}, F^{\text{EM}}_{tz})$ with $F^{\text{EM}}_{\mu\nu} = \partial_\mu A^{\text{EM}}_\nu - \partial_\nu A^{\text{EM}}_\mu$. 
The superscript ``EM'' distinguishes this Abelian gauge field from the non-Abelian QCD gauge field. 
By using Wick rotation $A^{\text{EM}}_0 \to iA^{\text{EM}}_4$, the Euclidean action becomes,
\begin{equation}
\begin{split}
&S_q = \int d^4 x^E \left( \bar{\psi} \sum_{j=1}^4 \gamma_j^E \partial_j \psi + \sum_{j=1}^4 \bar{\psi} i g \gamma_j^E A_j \psi \right.\\
&\left.- i Q_q e E_x x \, \bar{\psi} \gamma_4^E \psi \right),
\end{split}
\label{eq.2.1}
\end{equation}
Note that the replacement $A^{\text{EM}}_0 \to iA^{\text{EM}}_4$ effectively introduces an imaginary~(or Euclidean) electric field, which may also be interpreted as an analytic continuation.
By employing staggered fermions~\cite{ks,book2010}, the discretized action is,
\begin{equation}
\begin{split}
S_G &= \frac{\beta}{N_c} \sum_n \sum_{\mu>\nu} \operatorname{Re\,tr} \bigl[ 1 - U_{\mu\nu}(n) \bigr],\\
S_q &= \sum_n \left( \sum_{\mu} \sum_{\delta=\pm\mu} \bar{\chi}(n) \, U_{\delta}(n) V_{\delta}(n) \eta_{\delta}(n) \chi(n+\delta) \right.\\
&\left.+ 2am \bar{\chi}\chi \right),\\
\end{split}
\label{eq.2.2}
\end{equation}
where $a$ is the lattice spacing, $S_G$ is the Wilson gauge action~\cite{Wilsongauge,book2010} with $\beta = 2N_c/g_{\text{YM}}^2$, $g_{\text{YM}}$ being the gauge coupling, and $m$ denotes the fermion mass. The link variables are $U_\mu = e^{ia A_\mu}$ for QCD and $V_\mu = e^{ia e A^{\text{EM}}_\mu}$ for the electromagnetic field. The staggered phase factors are $\eta_\mu(n) = (-1)^{\sum_{\nu<\mu} n_\nu}$, and we use the conventions $U_{-\mu}(n) = U_\mu^\dagger(n-\mu)$, $V_{-\mu}(n) = V_\mu^*(n-\mu)$, $\eta_{-\mu}(n) = -\eta_\mu(n-\mu)$.
In this work, we use $N_f=1+1$~($u$ and $d$), and $am_{u,d}=0.05$.

To maintain $U(1)$ gauge invariance, the twisted boundary conditions~\cite{u1phaseisimaginary3,twist1,twist2,twist3} is introduced. 
For an electric field along $x$ in the axial gauge, $V_\mu(n)=1$ for all components except,
\begin{equation}
\begin{split}
&V_\tau(n) = e^{-i a Q_q e E_x (x - aL_x/2)},\\
&V_x\bigl(n_x = L_x-1\bigr) = e^{i a Q_q e E_x L_x (\tau - aL_{\tau}/2)},
\end{split}
\label{eq.2.3}
\end{equation}
with the electric field quantized as $a^2 e E_x = 6k\pi/(L_\tau L_x)$ so that $f$ takes the form $2k\pi/(L_\tau L_x)$ (here $|Q_q|=1/3$). 
In the actual simulation we employ a lattice of size $L_x \times L_y \times L_z \times L_\tau = 24 \times 24 \times 24 \times 6$.
Consequently, $a^2 e E_x = k \times \pi/24$ with $k=0,1,\dots,7$.
It should be noted that for $a^2 e E_x \sim \mathcal{O}(1)$ or larger, discretization errors become severe, therefore the upper limit of $E_x$ is chosen so that $a^2eE_x<1$.

\section{\label{sec:sec3}Numerical results}

To associate a lattice spacing and a temperature with each simulation parameter $\beta$, a matching procedure is performed on a lattice of size $L_x \times L_y \times L_z \times L_\tau = 24^3 \times 48$. 
The lattice spacing $a$ is extracted from the static quark potential $V(r)$~\cite{Bali:1992ab,Bali:2000vr,Orth:2005kq}. 
This is done by matching the calculated Sommer scale $r_0$ to its established physical value of $0.5\;\text{fm}$~\cite{Sommer:1993ce,Cheng:2007jq,MILC:2010hzw}.
The cases of $\beta = 5.3$ and $\beta=5.8$ are considered.
The resulting lattice spacings are~\cite{Yang:2026jma} $a^{-1}=770\pm 7\;{\rm MeV}$ and $2450\pm 2\;{\rm MeV}$, respectively.

\subsection{\label{sec:sec3.1}Chiral condensation and Polyakov loop}

The Polyakov loop and chiral condensate are related to the phase transition and have been studied in the presence of an external electric field in previous works.
Except for that, the charge density is also of interest in the study of quark mater in an external electric field.
In terms of staggered fermion field, they are~\cite{Bazavov:2011nk},
\begin{equation}
\begin{split}
&c_q(x)=\frac{1}{4L_yL_zL_{\tau}}\frac{2}{a^3} \sum_{n_x=a^{-1}x}\langle \bar{\chi }(n)\chi (n)\rangle,\\
&c_q=\frac{1}{L_x}\sum _{x}c_q(x),\\
&c^4_q(x)=\frac{1}{4L_yL_zL_{\tau}}\frac{1}{a^3} \\
&\times \sum_{n_x=a^{-1}x}\sum _{\delta = \pm }\eta _{\delta \tau}(n)\langle \bar{\chi }(n)U_{\delta\tau}(n)V_{\delta\tau}(n)\chi (n+\delta \hat{\tau})\rangle,\\
&P ({\bf n})=\frac{1}{N_c}\prod _{n_{\tau}}U_{\tau}({\bf n}, n_{\tau}) ,\\
&P (x) = \frac{1}{L_yL_z}\sum _{n_y,n_z} P({\bf n}=(a^{-1}x,n_y,n_z)),\\
\end{split}
\label{eq.3.1}
\end{equation}
where $\hat{\tau}$ denotes a dimensionless unit vector in $\tau$ direction.

At each $\beta$, we performed simulations of $TU_{th}+TU_m$ molecular dynamics time units~(TU) for each $k$.
At $k=0$, $TU_{th}+TU_m=500+1300$ TUs are simulated, then $k$ is increased, $TU_{th}+TU_m=200+1300$ TUs are simulated sequentially with increasing $k$.
For each $k$, the first $TU_{th}$ TUs are discarded for thermalization, and $TU_m$ configurations are measured.
Throughout this work, unless otherwise stated, statistical uncertainties are estimated using the formula $\sigma = \sqrt{2 \tau _{\rm int}} \sigma _{\rm jk}$~\cite{Gattringer:2010zz}, where $\tau _{\rm int}$ is the integrated autocorrelation time representing the number of TUs between statistically independent configurations, and $\sigma _{\rm jk}$ is the error obtained from a jackknife analysis. The autocorrelation time $\tau _{\rm int}$ is computed via an autocorrelation function analysis employing a window parameter $S 2$~\cite{Wolff:2003sm}.

\begin{figure*}
\begin{center}
\includegraphics[width=0.48\hsize]{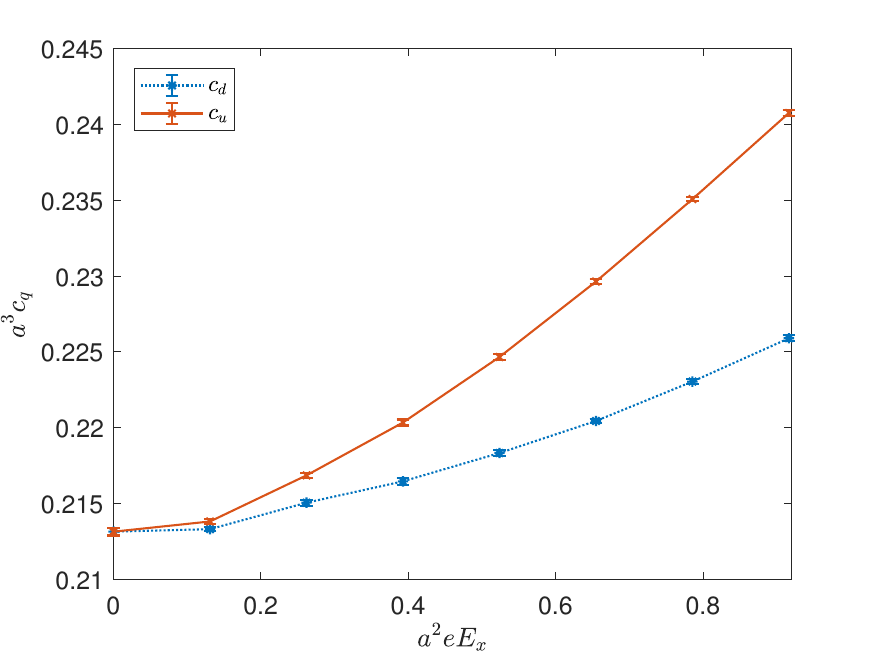}
\includegraphics[width=0.48\hsize]{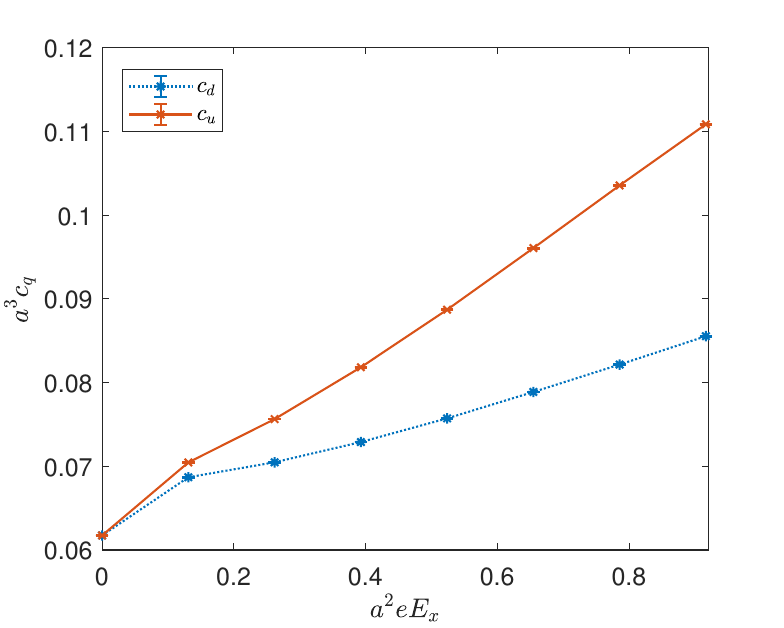}
\caption{\label{fig:chiral}$c_q$ as functions of $E_x$ at $\beta=5.3$~(left panel) and $\beta=5.8$~(right panel).}
\end{center}
\end{figure*}

$c_q$ as functions of $E_x$ at different $\beta$ are shown in Fig.~\ref{fig:chiral}.
We find that the chiral condensate is approximately a quadratic function of $E_x$, and the quadratic coefficient roughly satisfies that for the $u$ quark it is twice that for the $d$ quark, especially at larger $E_x$. 
Taking $\beta = 5.3$ as an example, fitting only the points with $k \ge 1$ gives the result $a^{3}c_u=0.214+0.015\left(a^2eE_x\right)^2$ and $a^{3}c_d=0.215+0.032\left(a^2eE_x\right)^2$. 
However, the $\chi^2/d.o.f.$ is too large, so this is only an approximate relation.
The case of $k=1$ is special, which indicates a discontinuity at non-zero imaginary electric field which has been discussed in Ref.~\cite{Endrodi:2023wwf}.
The behavior that the chiral condensate grows with imaginary electric field strength was also observed in Ref.~\cite{Yang:2023zzx}.

In the case of high temperature, the chiral condensation oscillates along $x$-axis exhibiting a forced density wave as a response of the external electric field.
It has been shown that the $c(x)$ can be fitted according to Ref.~\cite{Yang:2022zob},
\begin{equation}
\begin{split}
&a^3c_q(x)=A_q(E_x)+B_q(E_x)\cos \left(L_{\tau}Q_qaeE_x x\right).
\end{split}
\label{eq.3.2}
\end{equation}

\begin{table*}
\begin{center}
\begin{tabular}{c|ccc|ccc}
\hline    
$k$ & $A_u(E_x)$ & $B_u(E_x)$ & $\chi^2/d.o.f.$ & $A_d(E_x)$ & $B_d(E_x)$ & $\chi^2/d.o.f.$ \\
\hline
1 & $0.07045(4)$ & $-0.00501(6)$ & $2.82$ & $0.06868(4)$ & $0.00576(7)$ & $1.32$\\
2 & $0.07563(4)$ & $-0.00339(5)$ & $0.98$ & $0.07048(4)$ & $-0.00515(5)$ & $0.39$\\
3 & $0.08183(4)$ & $-0.00229(4)$ & $0.88$ & $0.07290(4)$ & $0.00431(4)$ & $0.37$\\
4 & $0.08872(4)$ & $-0.00173(4)$ & $0.42$ & $0.07573(4)$ & $-0.00339(4)$ & $0.66$\\
5 & $0.09607(5)$ & $-0.00158(5)$ & $0.27$ & $0.07889(4)$ & $0.00272(4)$ & $0.15$\\
6 & $0.10356(7)$ & $-0.00153(4)$ & $0.12$ & $0.08215(6)$ & $-0.00221(6)$ & $0.10$\\
7 & $0.11087(6)$ & $-0.00142(6)$ & $0.31$ & $0.08555(5)$ & $0.00192(5)$ & $0.29$\\
\hline
\end{tabular}
\end{center}
\caption{\label{tab.chiralfit}Results of fits of $c_q(x)$ according to Eq.~(\ref{eq.3.2}) at $\beta=5.8$.}
\end{table*}

\begin{figure*}
\begin{center}
\includegraphics[width=0.98\hsize]{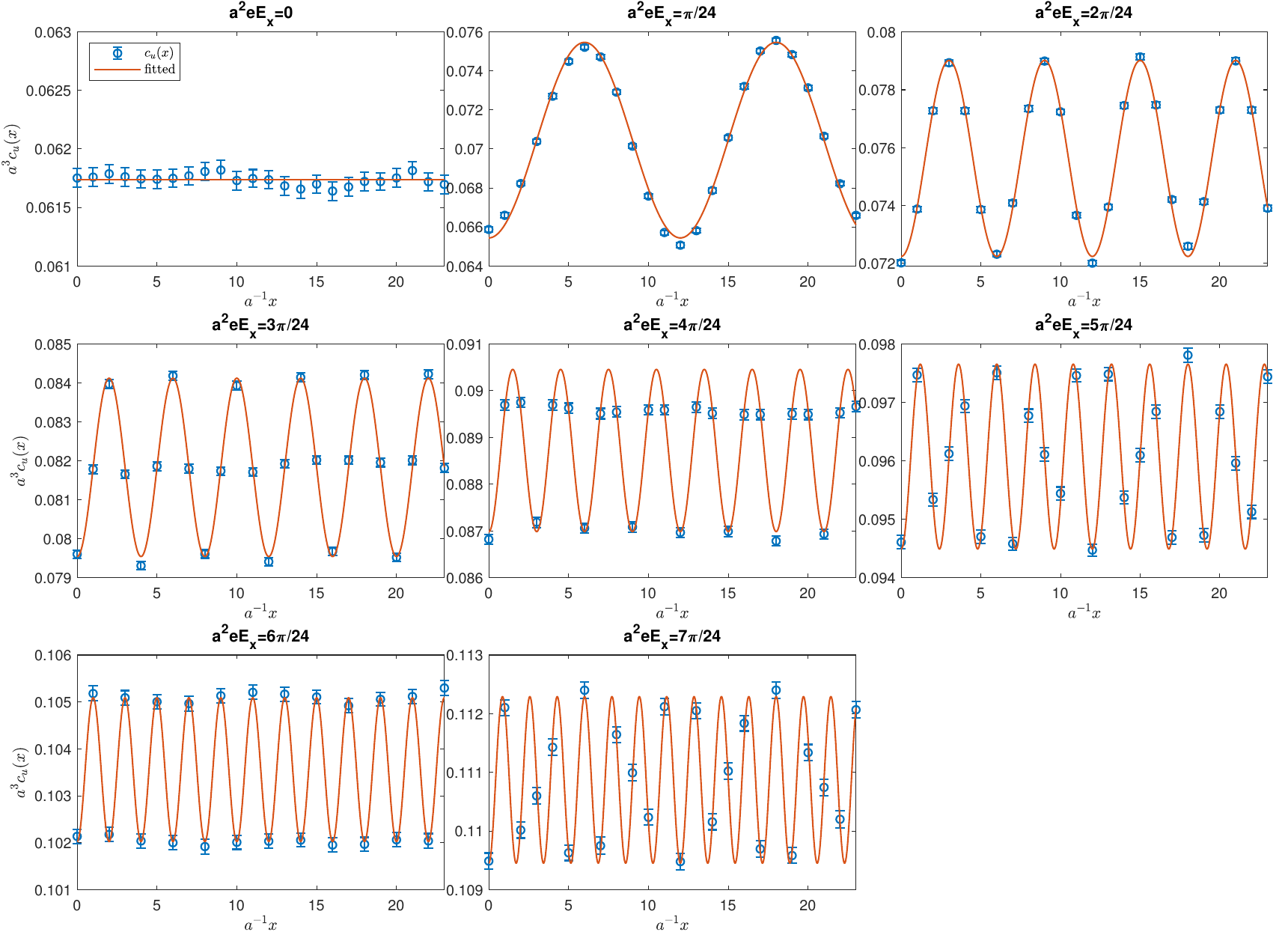}
\caption{\label{fig:chiralfit580u}The oscillations of $c_u(x)$ along $x$-axis and the fitted results according to Eq.~(\ref{eq.3.2}) at $\beta=5.8$.}
\end{center}
\end{figure*}

\begin{figure*}
\begin{center}
\includegraphics[width=0.98\hsize]{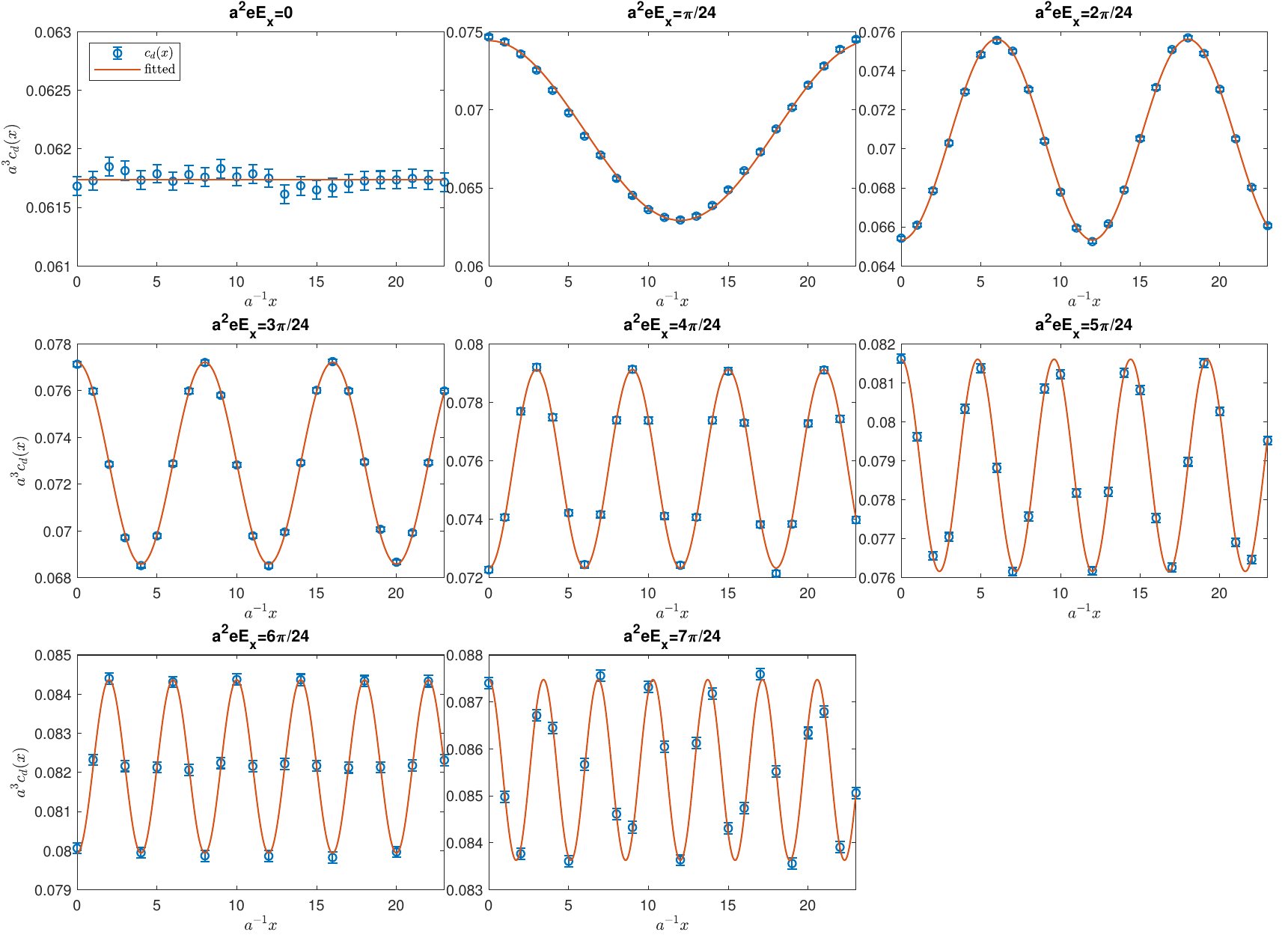}
\caption{\label{fig:chiralfit580d}Same as Fig.~\ref{fig:chiralfit580u} but for $c_d(x)$.}
\end{center}
\end{figure*}

A correlated fit is performed according to Eq.~(\ref{eq.3.2}). 
Autocorrelations between gauge configurations are approximately taken into account by rescaling the covariance matrix of the correlator by a factor $2\tau_{int}$, where $2\tau_{int}$ is the integrated autocorrelation time of $c_q(x)$. 
The maximum $\tau_{int}$ along the $c_q(x)$ is used.
Equivalently, this corresponds to rescaling the statistical uncertainties of the coefficients as $\sigma \to \sqrt{2\tau_{int}}\sigma$, and the goodness-of-fit as $\chi^2/d.o.f. \to \chi^2/d.o.f./(2\tau_{int})$.
In the following, the fits are always performed in this way.
The results of the fits are shown in Table~\ref{tab.chiralfit}, and Figs.~(\ref{fig:chiralfit580u}) and (\ref{fig:chiralfit580d}).
The spatial modulation of the chiral condensate same as Ref.~\cite{Yang:2022zob} is observed.

Except for the chiral condensate, it has been shown that, a non-trival imaginary charge distribution can be driven by the imaginary electric field at high temperatures~\cite{Endrodi:2021qxz,Endrodi:2022wym}, as~\cite{Yang:2022zob},
\begin{equation}
\begin{split}
&a^3{\rm Im}(c^4_q(x))=B^4_q(E_x)\sin \left(L_{\tau}Q_qaeE_x x\right).
\end{split}
\label{eq.3.2b}
\end{equation}

\begin{table}
\begin{center}
\begin{tabular}{c|cc|cc}
\hline    
$k$ & $B^4_u(E_x)$ & $\chi^2/d.o.f.$ & $B^4_d(E_x)$ & $\chi^2/d.o.f.$ \\
\hline
1 & $0.00919(6)$ & $5.93$ & $0.01007(5)$ & $3.24$\\
2 & $0.00690(3)$ & $5.93$ & $-0.00935(4)$ & $1.69$\\
3 & $0.00537(3)$ & $1.81$ & $0.00823(3)$ & $1.14$\\
4 & $0.00434(3)$ & $2.15$ & $-0.00690(3)$ & $2.89$\\
5 & $0.00446(3)$ & $1.45$ & $0.00587(2)$ & $0.89$\\
6 & $-$ & $-$ & $-0.00511(3)$ & $1.75$\\
7 & $0.00682(4)$ & $2.09$ & $0.00463(2)$ & $2.06$\\
\hline
\end{tabular}
\end{center}
\caption{\label{tab.chargefit}Results of fits of $c^4_q(x)$ according to Eq.~(\ref{eq.3.2b}) at $\beta=5.8$.
Note that for the case of $c^4_u$ and $k=6$, the ansatz becomes $B^4_u\sin (\pi n_x)$, which is zero, therefore the $B^4_u$ cannot be fitted.}
\end{table}

\begin{figure*}
\begin{center}
\includegraphics[width=0.98\hsize]{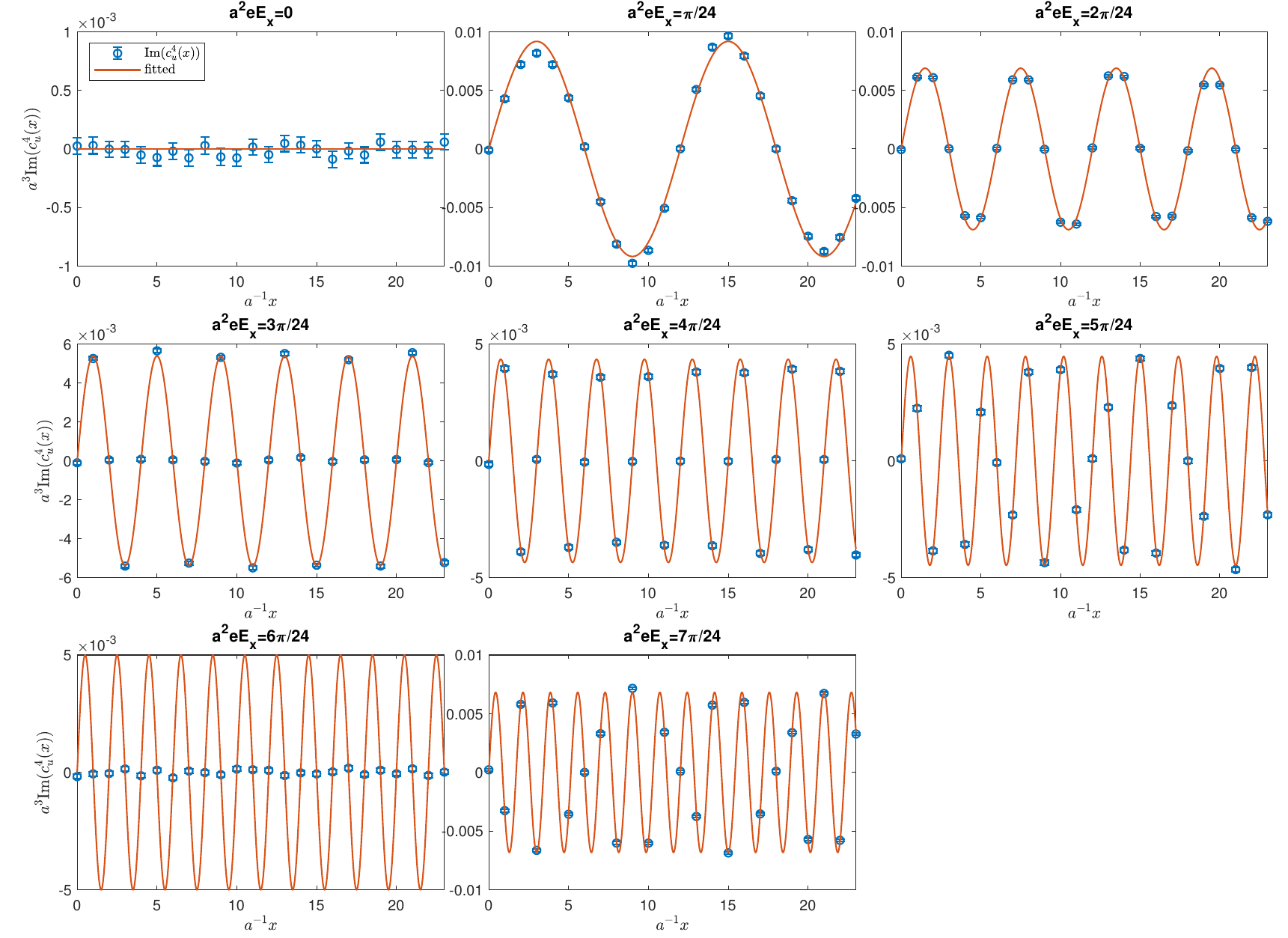}
\caption{\label{fig:c4fit580u}Same as Fig.~\ref{fig:chiralfit580u} but for $c^4_u(x)$.
Note that for the case of $c^4_u$ and $k=6$, $B^4_u$ cannot be fitted. 
Just for the sake of the schematic, we took an arbitrary $B^4_u$ in this special case.
We use a $B^4_u=0.005$ which is at the same order of $B^4_u$ in other cases.}
\end{center}
\end{figure*}

\begin{figure*}
\begin{center}
\includegraphics[width=0.98\hsize]{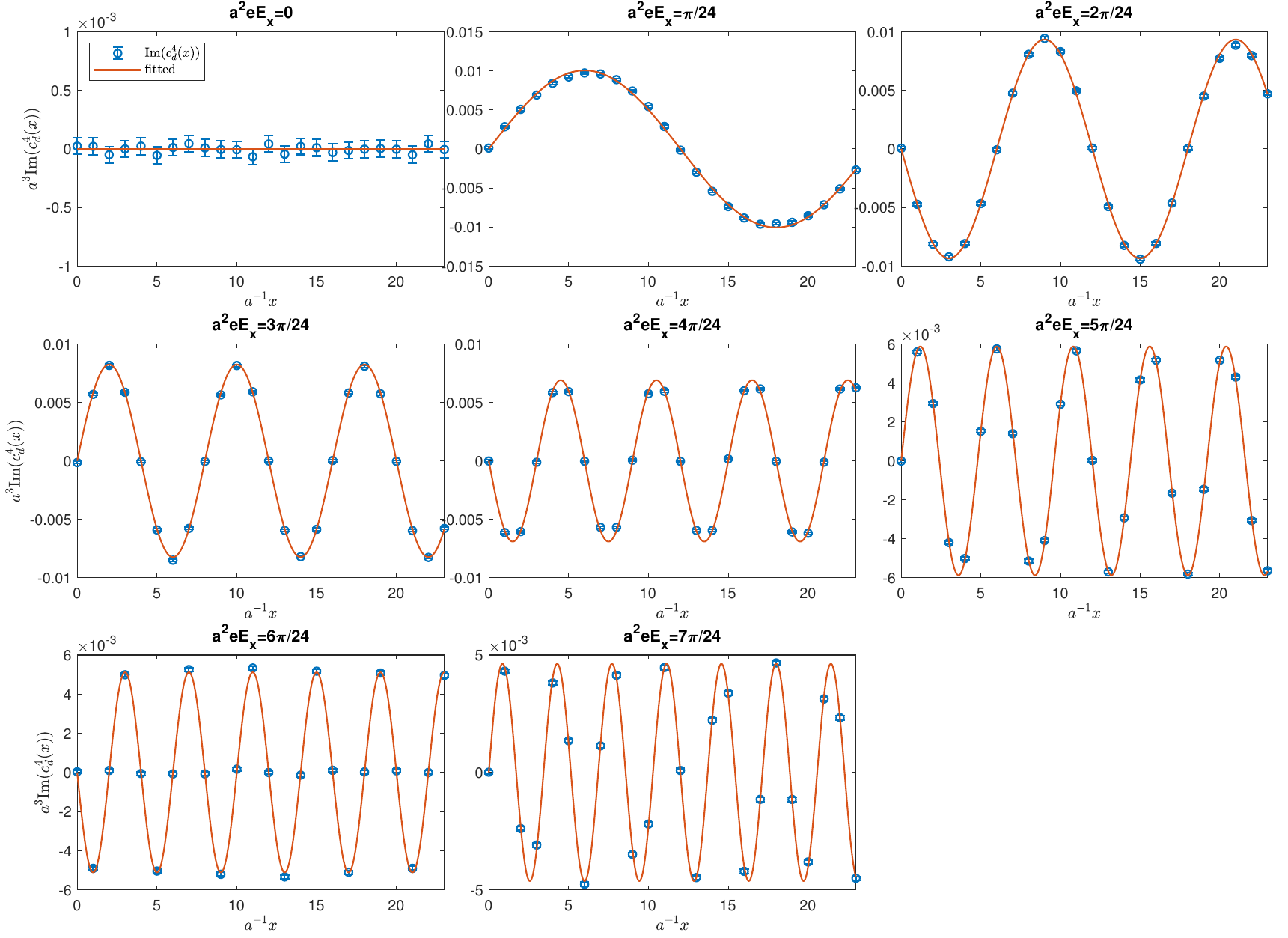}
\caption{\label{fig:c4fit580d}Same as Fig.~\ref{fig:chiralfit580u} but for $c^4_d(x)$.}
\end{center}
\end{figure*}

The fits of $c^4_q$ are shown in Figs.~\ref{fig:c4fit580u} and \ref{fig:c4fit580d}, and in Table~\ref{tab.chargefit}.
Note that for the case of $c^4_u$ and $k=6$, the ansatz becomes $B^4_u\sin (\pi n_x)$, which is zero, therefore the $B^4_u$ cannot be fitted.
Except for such a special case, the expected oscillating imaginary charge distribution can be reproduced.

According to Ref.~\cite{Yang:2022zob}, the Polyakov loop also exhibits spatially modulated behavior. 
Near the phase transition point,
\begin{equation}
\begin{split}
&P(x)=A_P+\sum _{q=u,d}B_{P,q}\exp \left(L_{\tau} i Q_qaeE_x x\right).\\
\end{split}
\label{eq.3.3}
\end{equation}
Therefore, when $A_P\ll B_{P,q}$, the winding number of $P(x)$ around the origin in the complex plane is proportional to $k$, while when $A_P\gg B_{P,q}$, the winding number is $0$, so that the range of ${\rm Im}(P(x))$ suddenly changes from $(-\pi,\pi)$ to approximately $(-\pi/2,\pi/2)$ at certain $T$ and $E_x$.
This behavior was observed at temperatures slightly above $T_c$ where $T_c$ is the pseudo-critical temperature when $k=0$.
In the cases we consider, the temperature is far from $T_c$, so the spatial modulation of $P(x)$ is not significant. 
Therefore, we only perform the fit for the two cases $k=1$ and $k=2$.
For the high-temperature case, Eq.~(\ref{eq.3.3}) no longer provides a good description. 
This is because the amplitude of the spatial modulation of the chiral condensate becomes significant at high temperatures, which inevitably affects $A_p$. 
Therefore, for the case of $\beta=5.8$, we also incorporate the effect of the chiral condensate into the ansatz and perform the fit using the following ansatz,
\begin{equation}
\begin{split}
P(x)&=A_P(E_x)+\sum _{q=u,d}B_{P,q}(E_x)\exp \left(L_{\tau} i Q_qaeE_x x\right)\\
&+\sum _{q=u,d}C_{P,q}(E_x)\cos \left(L_{\tau} Q_qaeE_x x\right).\\
\end{split}
\label{eq.3.4}
\end{equation}

\begin{figure*}
\begin{center}
\includegraphics[width=0.48\hsize]{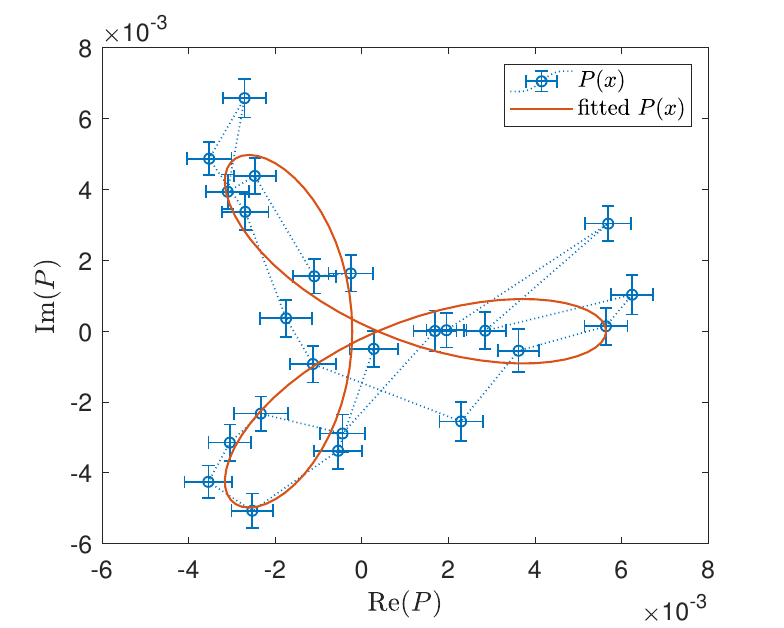}
\includegraphics[width=0.48\hsize]{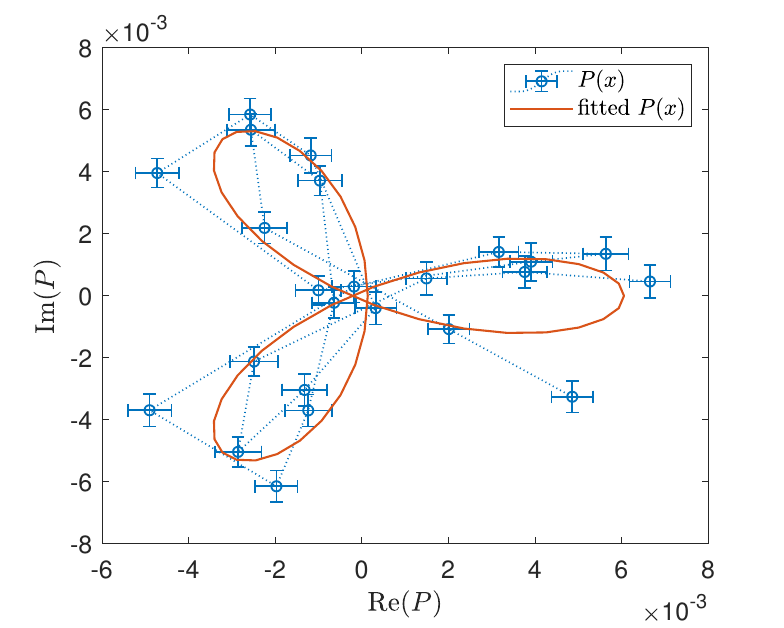}
\caption{\label{fig:pfit530}$P(x)$ along $x$-axis and the fitted results according to Eq.~(\ref{eq.3.3}) at $\beta=5.3$ in the case of $k=1$~(left panel) and $k=2$~(right panel).}
\end{center}
\end{figure*}
\begin{figure*}
\begin{center}
\includegraphics[width=0.48\hsize]{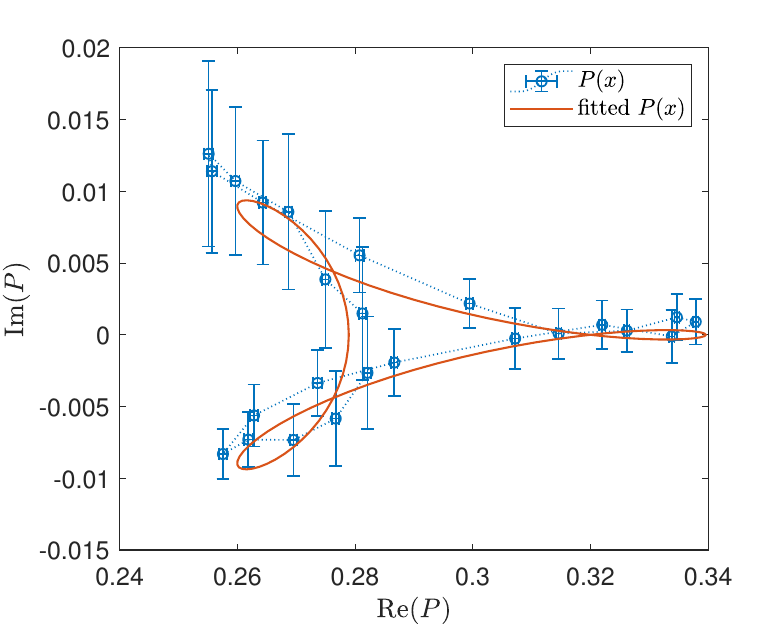}
\includegraphics[width=0.48\hsize]{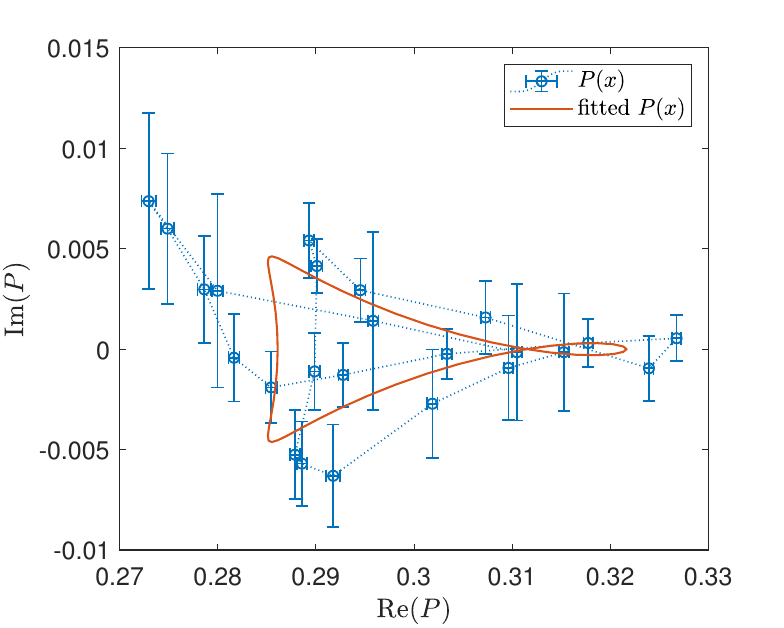}
\caption{\label{fig:pfit580}$P(x)$ along $x$-axis and the fitted results according to Eq.~(\ref{eq.3.4}) at $\beta=5.8$ in the case of $k=1$~(left panel) and $k=2$~(right panel).}
\end{center}
\end{figure*}

\begin{table*}
\begin{center}
\begin{tabular}{c|c|c|c|c|c|c}
\hline    
$k$ & $A_P(E_x)$ & $B_{P,u}(E_x)$ & $B_{P,d}(E_x)$ & $C_{P,u}(E_x)$ & $C_{P,d}(E_x)$ & $\chi^2/d.o.f.$ \\
\hline
1 & $-0.00003(20)$ & $-0.0029(2)$ & $0.0027(2)$ & & & $0.59$ \\
2 & $0.000007(187)$ & $-0.0030(2)$ & $0.0031(2)$ & & & $0.52$ \\
\hline
1 & $0.287(4)$ & $-0.007(3)$ & $0.004(2)$ & $-0.024(6)$ & $0.018(5)$ & $0.039$ \\
2 & $0.298(3)$ & $0.003(2)$  & $0.002(2)$ & $0.014(4)$  & $0.004(3)$ & $0.116$ \\
\hline
\end{tabular}
\end{center}
\caption{\label{tab.pfit}Results of fits of $P(x)$ according to Eqs.~(\ref{eq.3.3}) and (\ref{eq.3.4}) at $\beta=5.8$.}
\end{table*}

The fits are performed similar as $c_q(x)$ and $c^4_q(x)$ but with $\tau _{int}$ calculated by using $\left|P(x)\right|$.
The results are shown in Figs.~\ref{fig:pfit530} and \ref{fig:pfit530}, and in Table~\ref{tab.pfit}.
It can be seen from the results that the spatial modulation is also observed in our cases.

\subsection{\label{sec:sec3.2}The measurement of the mass of mesons}

The meson correlation function can be measured as,
\begin{equation}
\begin{split}
&C^{(i)}_{f_1f_2}(x)= \\
&\sum _{n_x=a^{-1}x,c_1,c_2} W_i(n) D^{-1}_{f_1}(n,c_1|0,c_2) \left((D^{\dagger})^{-1}_{f_2}\right)^*(n,c_1|0,c_2),\\
\end{split}
\label{eq.3.5}
\end{equation}
where $D^{-1}_{f_1}(n,c_1|0,c_2)$ denotes the matrix element of $D^{-1}$, $c_{1,2}$ are color indices, $W_i(n)$ is a sign function listed in Table~\ref{tab.mesonchannels} in the Appendix~\ref{sec:ap1}.
In this work, we employ local staggered meson operators, which are gauge invariant. 
Coulomb gauge fixing is nevertheless applied~\cite{Cucchieri:1996jm,Paciello:1991bd} in the computation of quark propagators for numerical convenience. 
Since the $du$ and $ud$ correlation functions yield identical results, we only consider the $uu$, $dd$, and $du$ channels.

With external electric field turned on, the correlation function does not necessarily satisfy the form in Eq.~(\ref{eq.ap.17}).
To investigate whether the exponential behavior still holds or not, the effective mass is introduced.

For this purpose, we can define,
\begin{equation}
\begin{split}
&\tilde{C}(n_x)=\frac{1}{2}\left(C^{(i)}_{f_1f_2}(n_x)+C^{(i)}_{f_1f_2}(L_x-n_x)\right),\\
&C^+(n_x)=\frac{1}{2}\left(\tilde{C}(n_x)+\tilde{C}(n_x+1)\right),\\
&C^-(n_x)=\frac{1}{2}(-1)^{n_x}\left(\tilde{C}(n_x)-\tilde{C}(n_x+1)\right),\\
&m_{eff}^{\pm}={\rm cosh}^{-1}\left(\frac{C^{\pm}(n_x+1)+C^{\pm}(n_x-1)}{2C^{\pm}(n_x)}\right),\\
\end{split}
\label{eq.3.6}
\end{equation}
when no confusion arises, the indices of $f_1,f_2$ and $(i)$ in the superscripts and subscripts are omitted.
Note that, the average over $n_x$ and $L_x-n_x$ in $\tilde{C}(n_x)$ does not apply for $n_x=0$ and $L_x/2$ points.
Since $C^{(i)}_{f_1f_2}(n_x)=C^+(n_x)+(-1)^{n_x}C^-(n_x)$, so when $m_-\ll m_+$, $C^+(n_x)$ approximately extract the correlation function according to $m_+$, and when $m_+\ll m_-$, $C^-(n_x)$ approximately extract the correlation function according to $m_-$.
In this work, a clear dominance of either positive or negative parity is observed in $\tilde{C}(x)$, in the following study we consider only one channel and perform the fit using the ansatz with only one parity,
\begin{equation}
\begin{split}
&\tilde{C}^{\pm}(n_x)=(\pm)^{n_x}A_m\left(e^{-m^{\pm}n_x}+e^{-m^{\pm}(L_x-n_x)}\right).\\
\end{split}
\label{eq.3.7}
\end{equation}
In the following, we directly use the particle symbols for positive parity or negative parity given in Table~\ref{tab.mesonchannels} to denote the corresponding channels.

\subsubsection{\label{sec:sec3.3.1}The case at lower temperature}

\begin{figure*}
\begin{center}
\includegraphics[width=0.32\hsize]{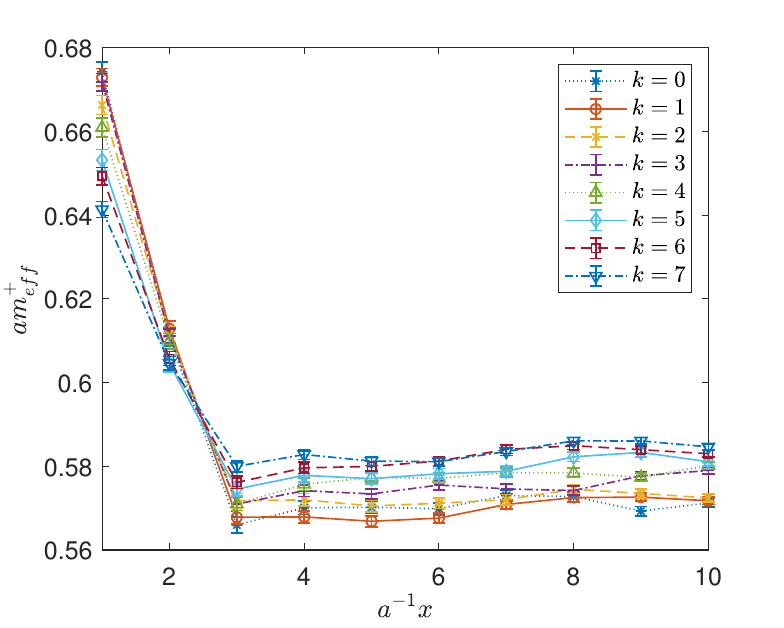}
\includegraphics[width=0.32\hsize]{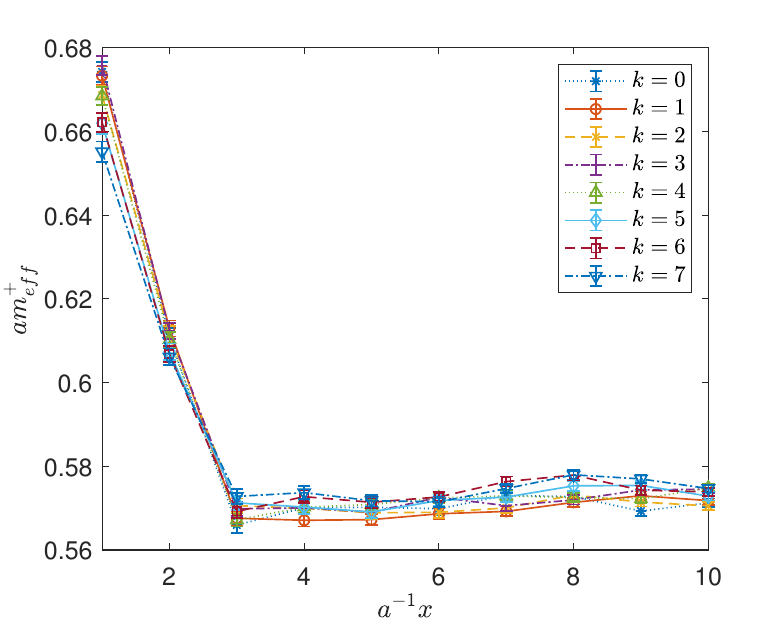}
\includegraphics[width=0.32\hsize]{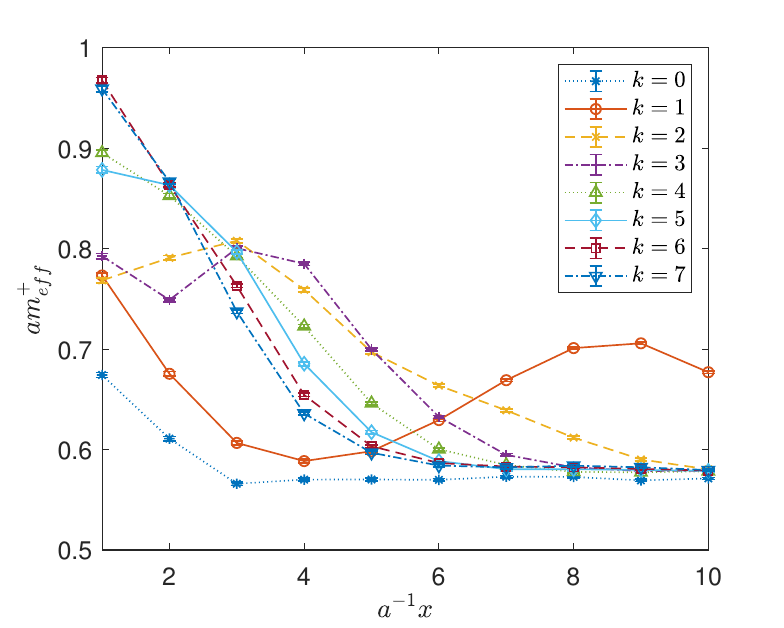}
\caption{\label{fig:effmass5301}Effective masses of $a_0$ channels at $\beta=5.3$. The results of $uu$ channel at different $k$ are shown in the left panel, the results of $dd$ channel are shown in the middle panel, and the results of $du$ channel are shown in the right panel.}
\end{center}
\end{figure*}

The effective masses of $a_0$ channels at different external electric field strengths are shown in Fig.~\ref{fig:effmass5301}.
From the figure, it can be seen that the $uu$ and $dd$ channels exhibit clear plateaus. 
Moreover, the $uu$ channel shows that the meson mass increases with the strength of the external electric field. 
However, the $du$ channel displays different behavior. 
From the $k = 1$ result, we observe a suspected spatial modulation in the $du$ channel, but this phenomenon disappears for larger $k$.
At larger $E_x$, the position-dependent phase varies more rapidly, possibly leading to loss of resolution, which may reduce the visibility of the oscillatory signal. 

\begin{table}
\begin{center}
\begin{tabular}{c|cc|cc|cc}
\hline    
$k$ & $am_{uu}$  & $\chi^2/d.o.f.$ & $am_{dd}$ & $\chi^2/d.o.f.$ & $am_{du}$ & $\chi^2/d.o.f.$ \\
\hline
0 & $0.5720(3)$ & $1.93$ & & & \\
1 & $0.5714(4)$ & $3.16$ & $0.5709(4)$ & $2.29$ & $0.6769(14)$ & $185.60$ \\
2 & $0.5734(3)$ & $3.17$ & $0.5716(3)$ & $3.51$ & $0.5936(9)$ & $247.63$ \\
3 & $0.5766(3)$ & $3.58$ & $0.5729(4)$ & $2.79$ & $0.5821(5)$ & $13.91$ \\
4 & $0.5785(3)$ & $2.05$ & $0.5730(3)$ & $1.52$ & $0.5793(3)$ & $3.56$ \\
5 & $0.5807(3)$ & $4.12$ & $0.5733(3)$ & $3.22$ & $0.5798(3)$ & $0.25$ \\
6 & $0.5836(3)$ & $3.74$ & $0.5753(3)$ & $3.69$ & $0.5816(3)$ & $1.42$ \\
7 & $0.5845(3)$ & $4.03$ & $0.5755(3)$ & $4.65$ & $0.5820(3)$ & $1.26$ \\
\hline
\end{tabular}
\end{center}
\caption{\label{tab.massfit1}Fitted masses of $a_0$ channels at $\beta=5.3$.}
\end{table}

The results of fits for $a_0$ are shown in Table~\ref{tab.massfit1}. 
For the $uu$ and $dd$ channels, we perform the fit in the range $3 \leq n_x \leq L_x/2$. 
For the $du$ channel, we ignore the spatial oscillation and perform the fit in the range $6 \leq n_x \leq L_x/2$. 
It can be seen from the table that the masses of the $uu$ and $dd$ channels indeed increase with the external electric field. 
For the $du$ channel, the $\chi^2/d.o.f.$ is very large, which is due to the absence of a well-formed plateau.

\begin{figure*}
\begin{center}
\includegraphics[width=0.32\hsize]{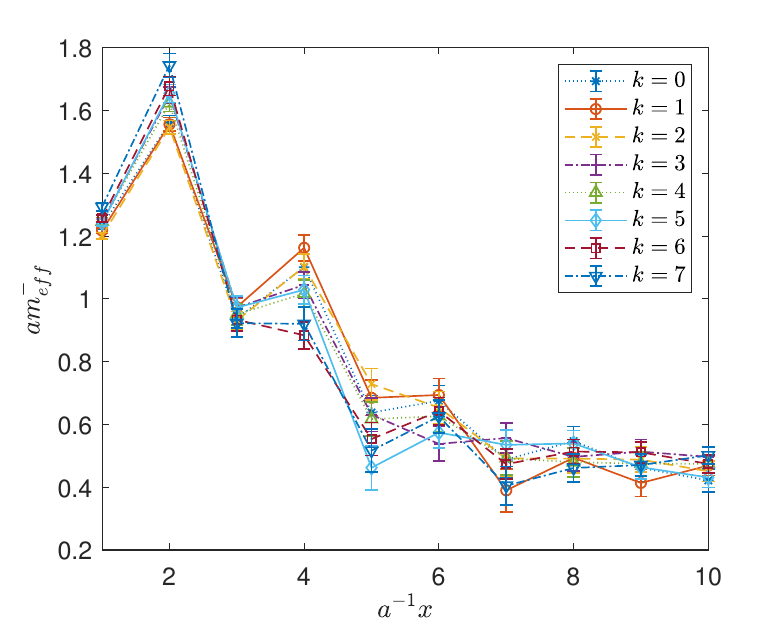}
\includegraphics[width=0.32\hsize]{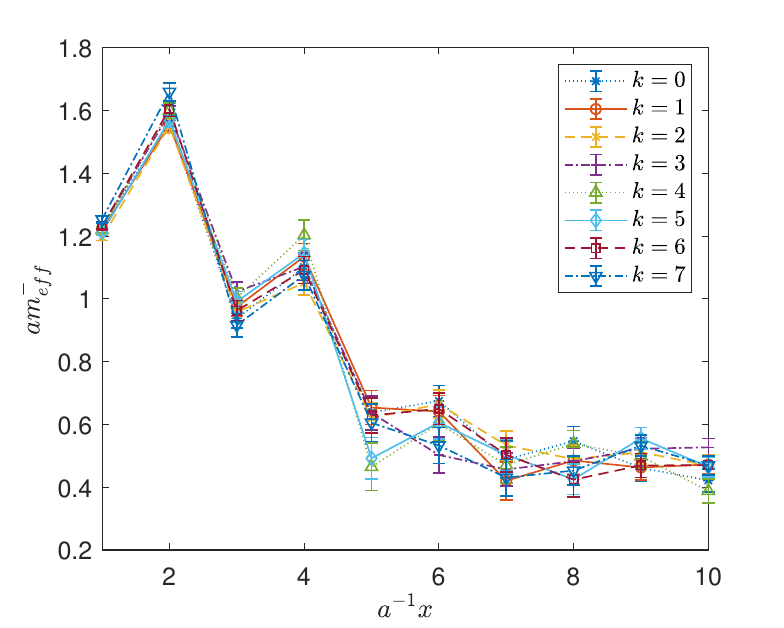}
\includegraphics[width=0.32\hsize]{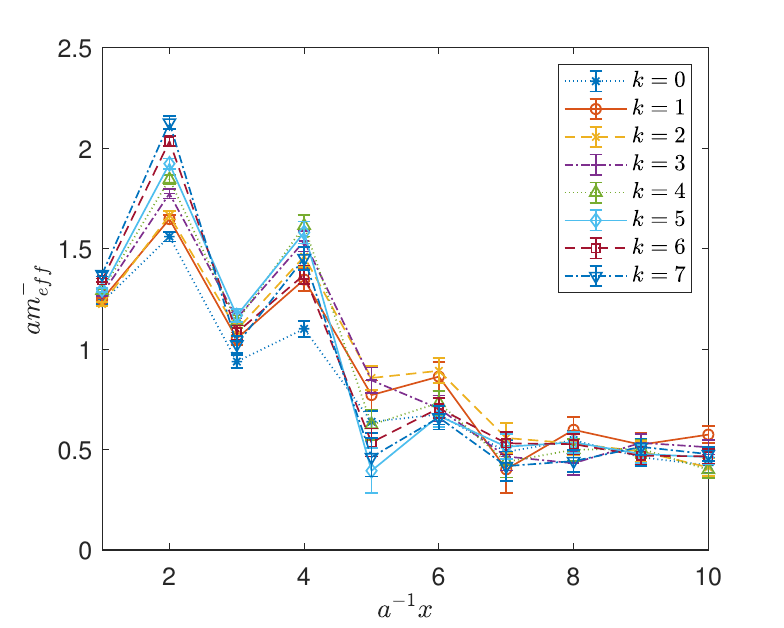}
\caption{\label{fig:effmass5302}Same as Fig.~\ref{fig:effmass5301} but for $\pi$ channels at $\beta=5.3$.}
\end{center}
\end{figure*}

The effective masses of $\pi$ channels at different external electric field strengths are shown in Fig.~\ref{fig:effmass5302}.
In the pseudo-scalar channel, the correlators exhibit a clear oscillatory behavior in $x$, which can be attributed to the opposite-parity contributions inherent in the staggered formulation. 
A plateau is observed for $x \geq 5$, indicating the dominance of the lowest screening state at sufficiently large separation. 
Within uncertainties, the extracted effective masses show little dependence on the external electric field. 

\begin{table}
\begin{center}
\begin{tabular}{c|cc|cc|cc}
\hline    
$k$ & $am_{uu}$  & $\chi^2/d.o.f.$ & $am_{dd}$ & $\chi^2/d.o.f.$ & $am_{du}$ & $\chi^2/d.o.f.$ \\
\hline
0 & $0.50(3)$ & $0.48$ & & & \\
1 & $0.49(2)$ & $0.80$ & $0.48(2)$ & $0.35$ & $0.56(4)$ & $0.65$ \\
2 & $0.48(2)$ & $0.74$ & $0.50(2)$ & $0.80$ & $0.48(3)$ & $2.00$ \\
3 & $0.50(2)$ & $0.10$ & $0.49(2)$ & $0.17$ & $0.50(2)$ & $0.39$ \\
4 & $0.49(2)$ & $0.41$ & $0.49(2)$ & $0.52$ & $0.49(2)$ & $0.63$ \\
5 & $0.50(2)$ & $0.45$ & $0.50(2)$ & $0.63$ & $0.50(2)$ & $0.61$ \\
6 & $0.51(2)$ & $0.42$ & $0.48(2)$ & $0.47$ & $0.51(2)$ & $0.64$ \\
7 & $0.49(2)$ & $0.42$ & $0.48(2)$ & $0.21$ & $0.49(2)$ & $0.33$ \\
\hline
\end{tabular}
\end{center}
\caption{\label{tab.massfit2}Same as Table~\ref{tab.massfit1} but for the $\pi$ channel.}
\end{table}

The results of fits for $\pi$ are shown in Table~\ref{tab.massfit2}. 
For all channels, we perform the fit in the range $5 \leq n_x \leq L_x/2$. 
It can be seen from the table that the masses of the all channels indeed are insensitive to the external electric field, except for the case of $k=1$ in the $du$ channel, where the mass is about $10\%$ larger. 
The origin of this deviation is not entirely clear. 

\begin{figure*}
\begin{center}
\includegraphics[width=0.32\hsize]{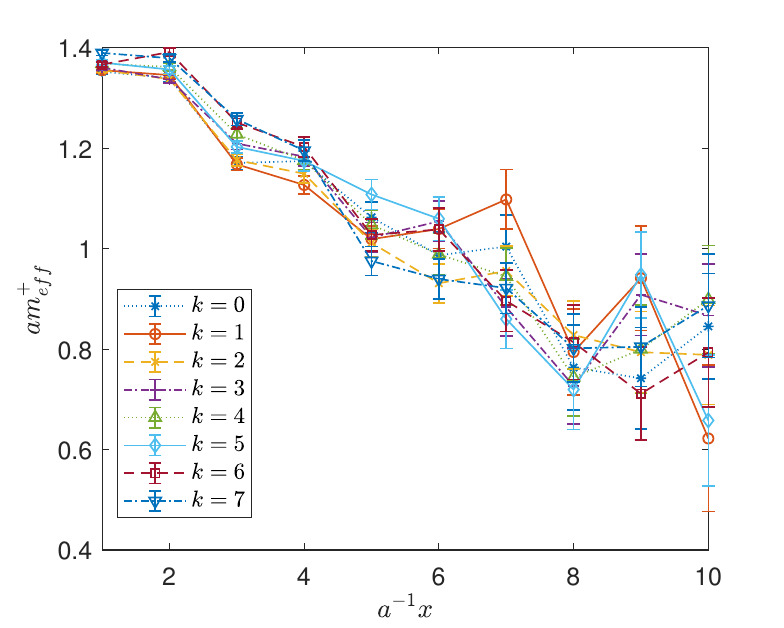}
\includegraphics[width=0.32\hsize]{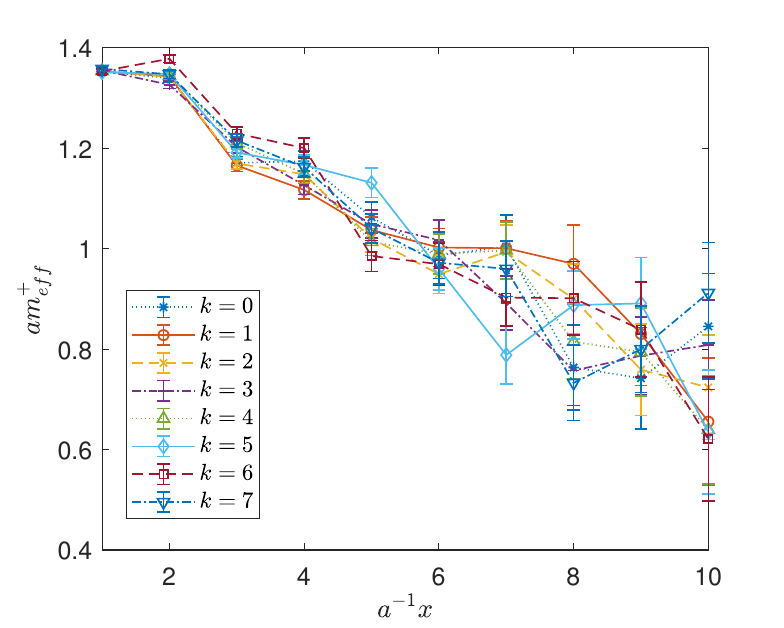}
\includegraphics[width=0.32\hsize]{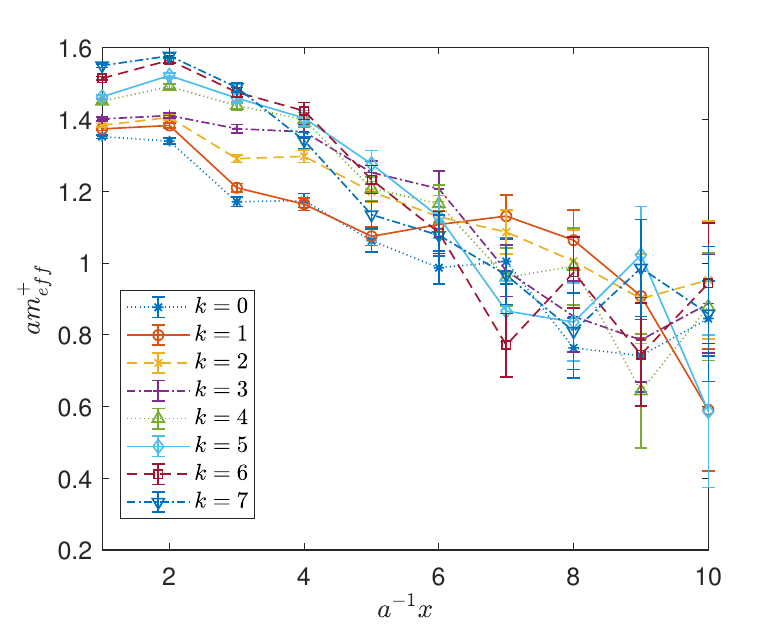}
\caption{\label{fig:effmass5303}Same as Fig.~\ref{fig:effmass5301} but for $a_{1,\perp}$ channels at $\beta=5.3$.}
\end{center}
\end{figure*}

\begin{table}
\begin{center}
\begin{tabular}{c|cc|cc|cc}
\hline    
$k$ & $am_{uu}$  & $\chi^2/d.o.f.$ & $am_{dd}$ & $\chi^2/d.o.f.$ & $am_{du}$ & $\chi^2/d.o.f.$ \\
\hline
0 & $0.83(4)$ & $1.02$ & & & \\
1 & $0.93(6)$ & $1.47$ & $0.89(5)$ & $1.06$ & $1.03(5)$ & $0.90$ \\
2 & $0.86(3)$ & $0.58$ & $0.86(4)$ & $1.49$ & $1.01(5)$ & $0.35$ \\
3 & $0.82(3)$ & $0.59$ & $0.79(3)$ & $0.48$ & $0.87(5)$ & $0.49$ \\
4 & $0.85(3)$ & $0.60$ & $0.84(4)$ & $1.27$ & $0.86(5)$ & $0.76$ \\
5 & $0.85(3)$ & $1.10$ & $0.83(3)$ & $0.37$ & $0.86(4)$ & $0.40$ \\
6 & $0.84(3)$ & $0.35$ & $0.86(4)$ & $0.39$ & $0.85(4)$ & $0.25$ \\
7 & $0.86(4)$ & $0.21$ & $0.85(4)$ & $0.52$ & $0.91(6)$ & $0.13$ \\
\hline
\end{tabular}
\end{center}
\caption{\label{tab.massfit3}Same as Table~\ref{tab.massfit1} but for the $a_{1,\perp}$ channel.}
\end{table}

The effective masses of $a_{1,\perp}$ channels at different external electric field strengths are shown in Fig.~\ref{fig:effmass5303}.
In the axial-vector channel, the effective masses exhibit relatively large statistical uncertainties, and no clear plateau is observed. 
This suggests that the correlators are significantly affected by excited-state contributions and that the ground-state dominance is not well established. 
Within the present accuracy, no significant dependence of the extracted masses on the external electric field is observed. 
For sufficiently large $x$ and within a sufficiently small interval, a plateau can be considered to approximately emerge within the uncertainties. 
Therefore, we perform the fit for $6 \leq n_x \leq L_x/2$. 
The results are shown in Table~\ref{tab.massfit3}.

\begin{figure}
\begin{center}
\includegraphics[width=0.48\hsize]{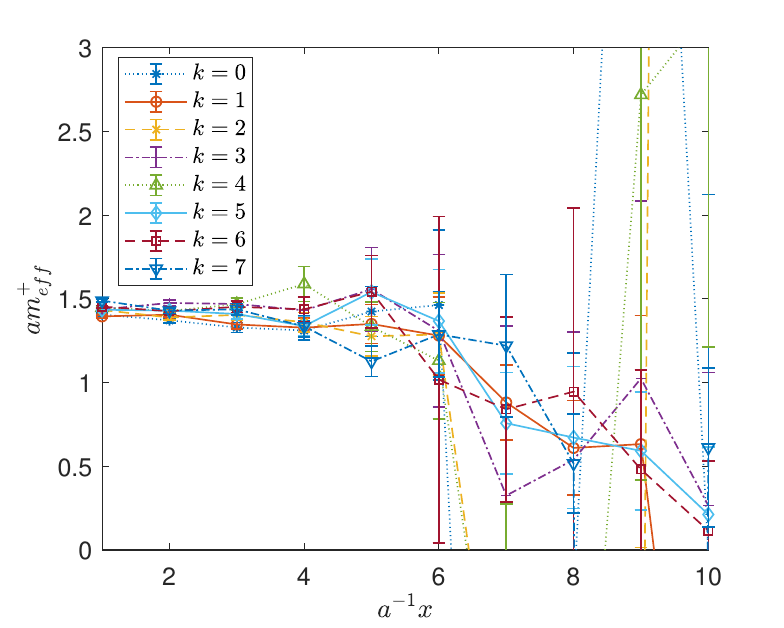}
\includegraphics[width=0.48\hsize]{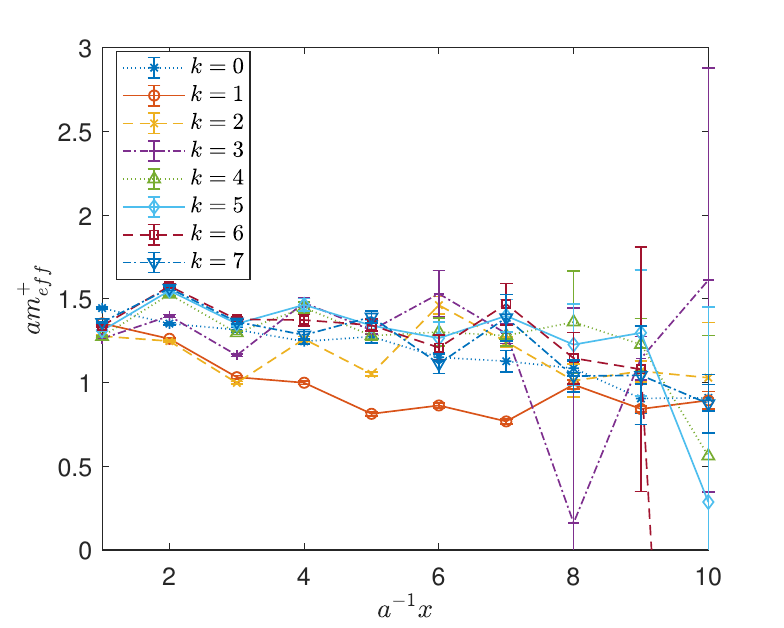}
\caption{\label{fig:effectivemass4}The effective masses of the temporal axial vector in the $uu$ channel at $\beta=5.3$~(left panel) and $\beta=5.8$~(right panel).}
\end{center}
\end{figure}
For the other channels, the errors of the correlation functions are too large, so we do not perform fits. 
As an example, the effective masses of the temporal axial vector are shown in Fig.~\ref{fig:effectivemass4}, where it can be seen that the plateau is no longer identifiable.

\subsubsection{\label{sec:sec3.3.2}The case at higher temperature}

\begin{figure*}
\begin{center}
\includegraphics[width=0.32\hsize]{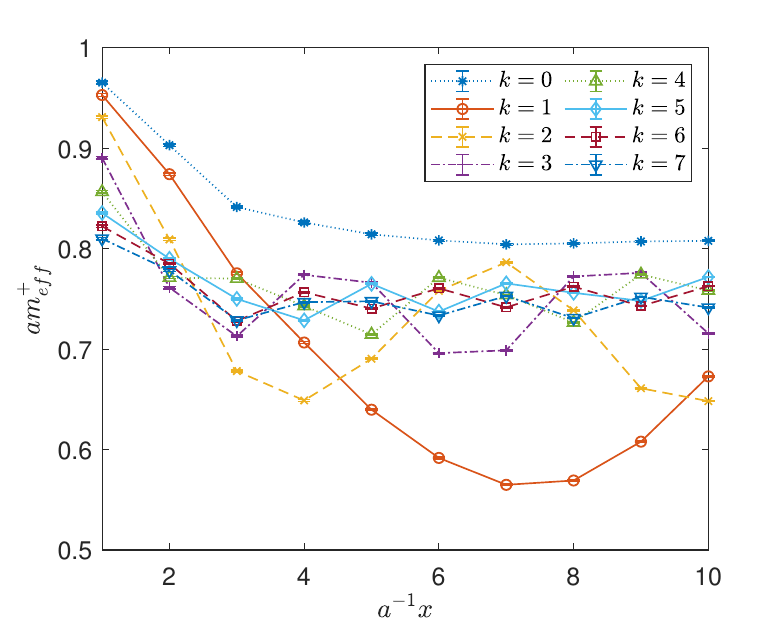}
\includegraphics[width=0.32\hsize]{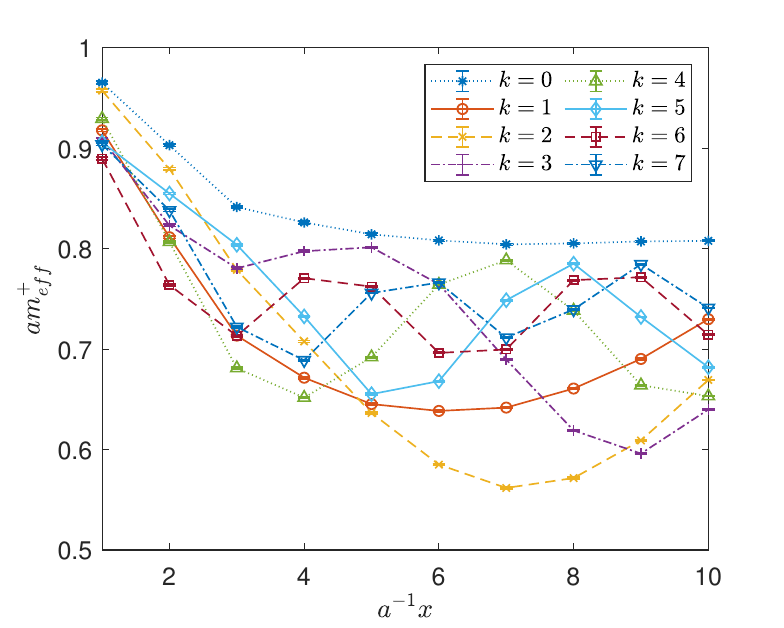}
\includegraphics[width=0.32\hsize]{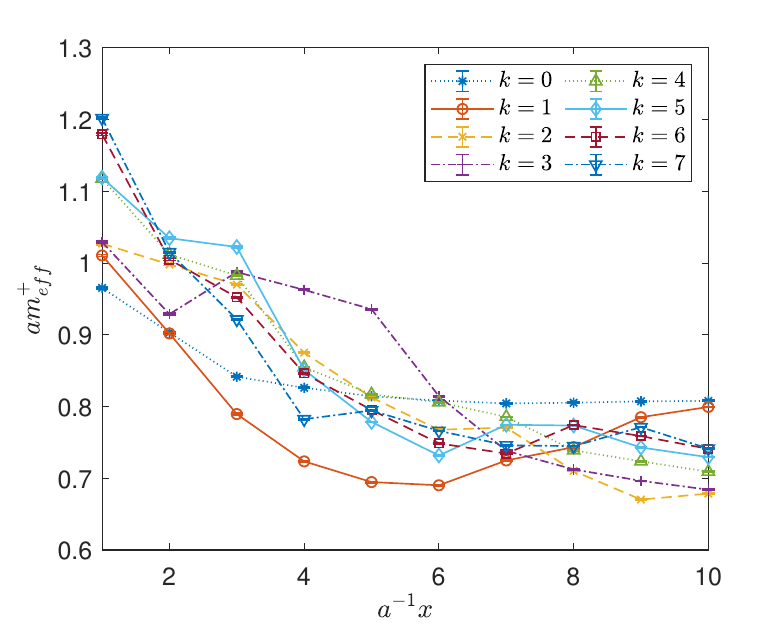}
\caption{\label{fig:effmass5801}Same as Fig.~\ref{fig:effmass5301} but for $a_0$ channels at $\beta=5.3$.}
\end{center}
\end{figure*}

The effective masses of $a_0$ channels at high temperature at different external electric field strengths are shown in Fig.~\ref{fig:effmass5801}.
From Fig.~\ref{fig:effmass5801}, we can clearly see the oscillation pattern of $m_{eff}$ in the $uu$ and $dd$ channels as functions of $x$. 
It can be observed that the $uu$ and $dd$ channels follow the same behavior, with an oscillation frequency of $a L_{\tau} Q_q eE_x$. 
These frequencies are consistent with the spatial oscillation frequencies of the chiral condensates and charge densities $c_{u,d}(x)$.

To analyze the oscillatory behavior, we perform a discrete Fourier transform of the effective mass,
\begin{equation}
\begin{split}
&m_p(p)=\frac{1}{\sqrt{n}}\sum _{a^{-1}x=1}^{n}m_{eff}(x)e^{\frac{2\pi i}{n}(a^{-1}x-1)ap},\\
\end{split}
\label{eq.3.8}
\end{equation}
where $p$ is the lattice momentum, $n=10$.
Since the effective mass does not exhibit a clear plateau and contains a slowly varying component, the Fourier spectrum is dominated by low-momentum contributions. 
As a result, only sufficiently strong and coherent oscillatory modes can be identified as distinguishable features above this background.

\begin{figure}
\begin{center}
\includegraphics[width=0.48\hsize]{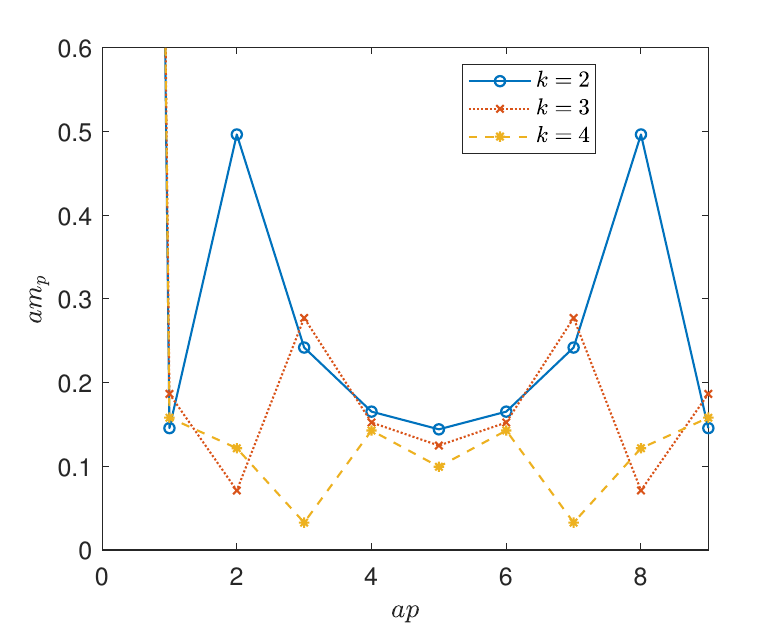}
\includegraphics[width=0.48\hsize]{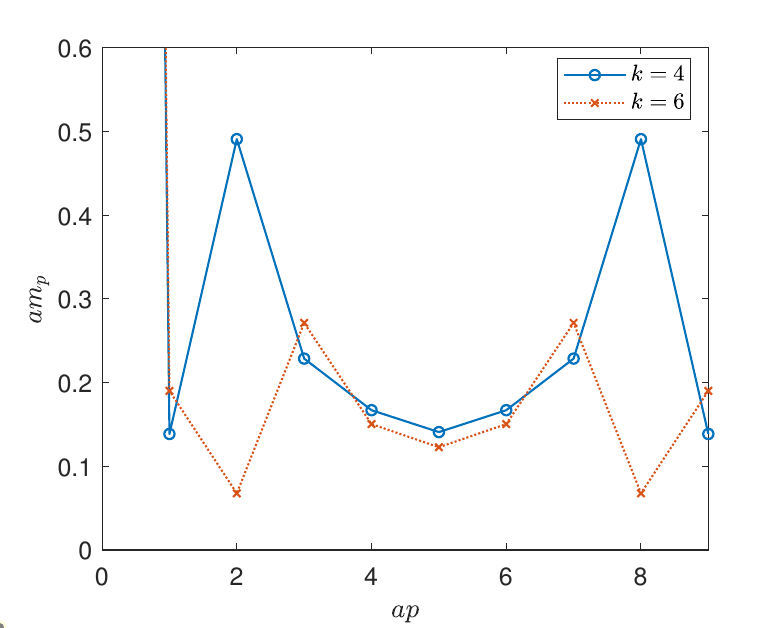}
\caption{\label{fig:mk580uu}$m_p$ defined in Eq.~(\ref{eq.3.8}) for the $uu$~(left panel) and $dd$~(right panel) channels at different $k$~(electric field strength).
The corresponding structure at $n-ap$, is related to the one at $ap$ by the discrete Fourier symmetry, contains no additional information and will be ignored.}
\end{center}
\end{figure}
In the discrete Fourier transform, only the center values of $m_{eff}$ are used.
The results of $m_p$ for $uu$ and $dd$ channels of $a_0$ are shown in Fig.~\ref{fig:mk580uu}.
In the single-flavor channels, identifiable peaks remain visible, indicating the presence of a dominant oscillatory component. 
In contrast, for the mixed-flavor channel, no distinct peak can be resolved, therefore the case of $du$ channel is not shown.
The absence of peaks suggesting that the oscillatory behavior is either weaker or hidden by the low-momentum background.
Due to the absence of clear plateaus in the effective masses, especially in the presence of strong oscillations, we do not attempt to extract screening masses from fits in this regime and instead focus on the behavior of $m_{eff}$.

\begin{figure*}
\begin{center}
\includegraphics[width=0.32\hsize]{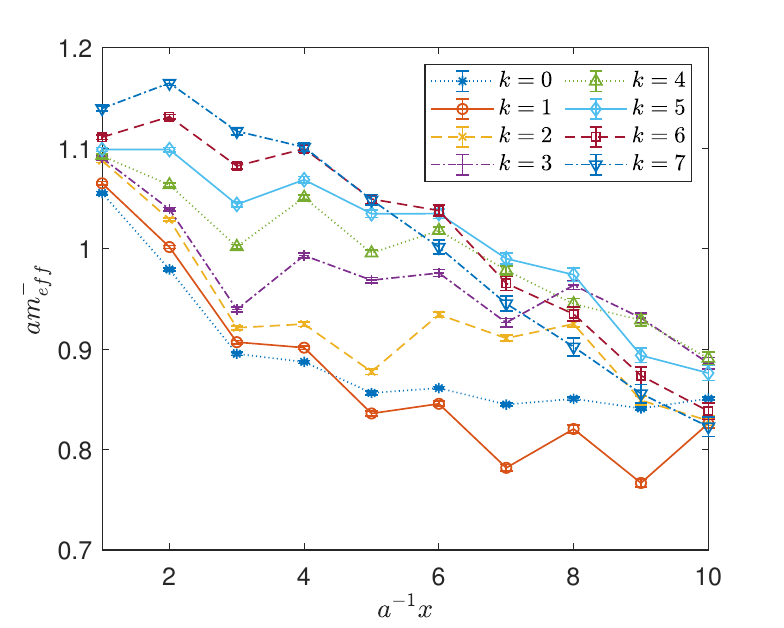}
\includegraphics[width=0.32\hsize]{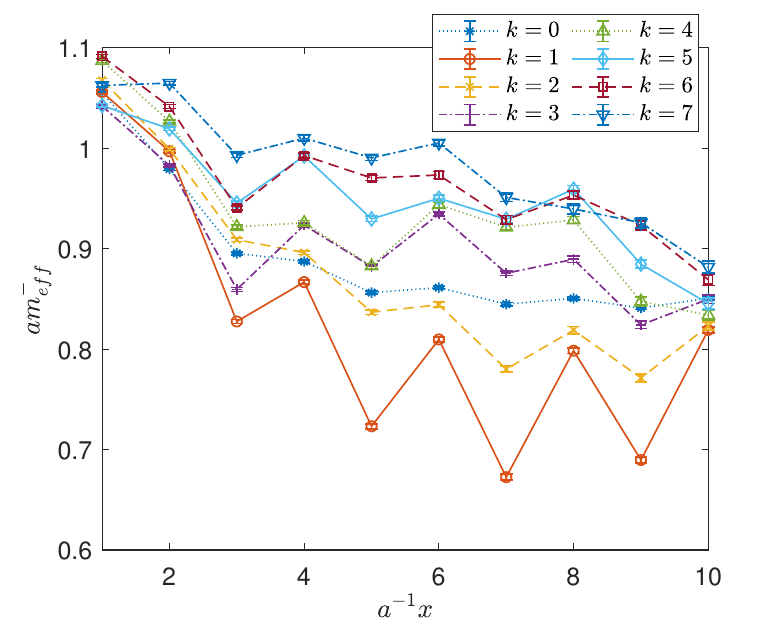}
\includegraphics[width=0.32\hsize]{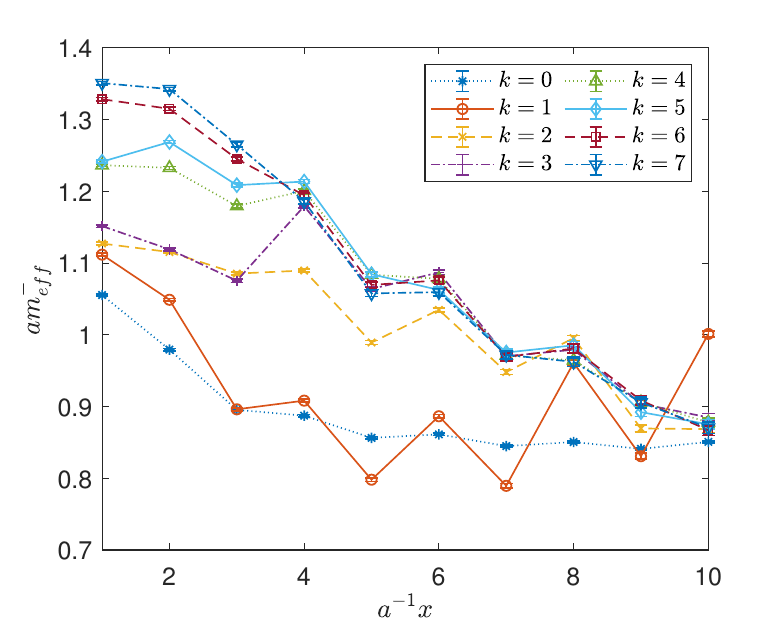}
\caption{\label{fig:effmass5802}Same as Fig.~\ref{fig:effmass5301} but for $\pi$ channels at $\beta=5.8$.}
\end{center}
\end{figure*}

\begin{figure*}
\begin{center}
\includegraphics[width=0.32\hsize]{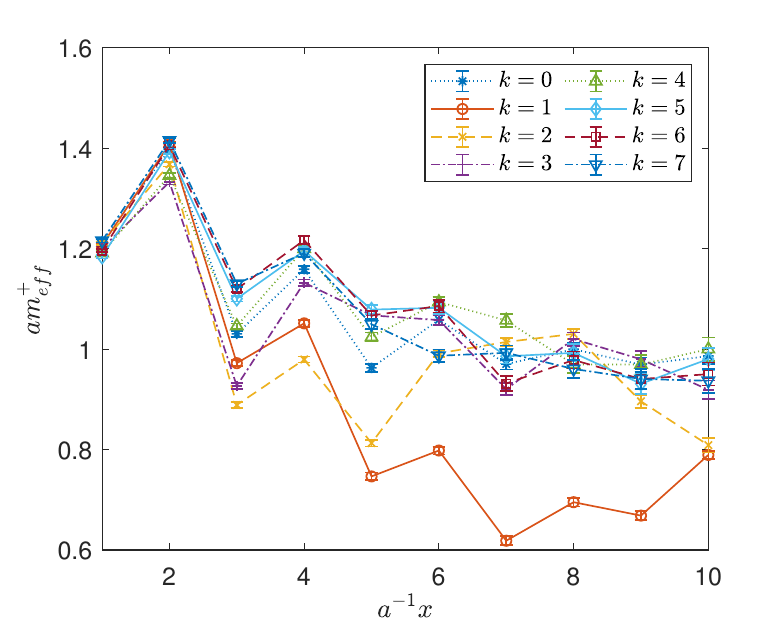}
\includegraphics[width=0.32\hsize]{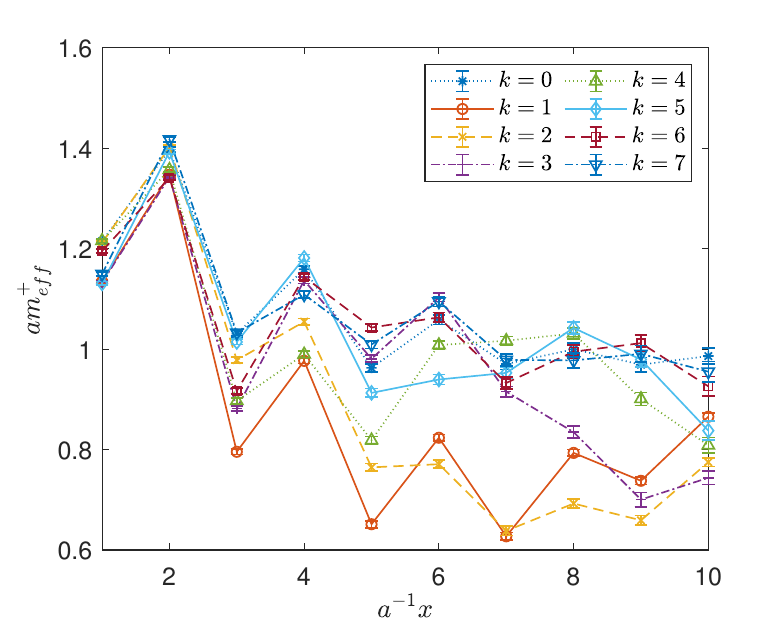}
\includegraphics[width=0.32\hsize]{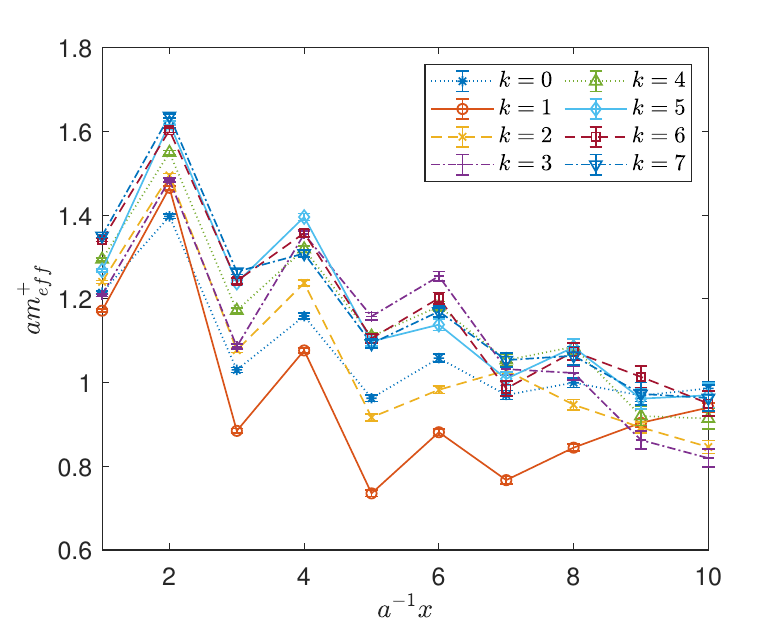}
\caption{\label{fig:effmass5803}Same as Fig.~\ref{fig:effmass5301} but for $a_{1,\perp}$ channels at $\beta=5.8$.}
\end{center}
\end{figure*}

The effective masses of $\pi$ and $a_{1,\perp}$ channels at high temperature and at different external electric field strengths are shown in Figs.~\ref{fig:effmass5802} and \ref{fig:effmass5803}, respectively.
In the pseudo-scalar and axial-vector channels at high temperature, the oscillatory behavior induced by the external electric field is less evident in the effective masses due to the intrinsic $(-1)^{n_x}$ oscillation associated with the staggered formulation.
Nevertheless, a residual modulation can still be identified in the $uu$ channel for small field strengths~($k=1$ and $k=2$), indicating that the same underlying mechanism is at work. 

\begin{figure}
\begin{center}
\includegraphics[width=0.48\hsize]{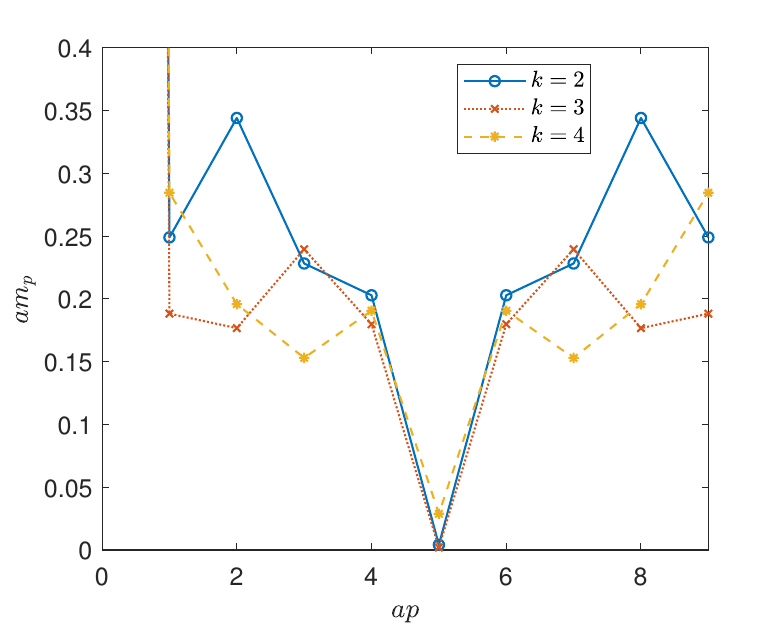}
\includegraphics[width=0.48\hsize]{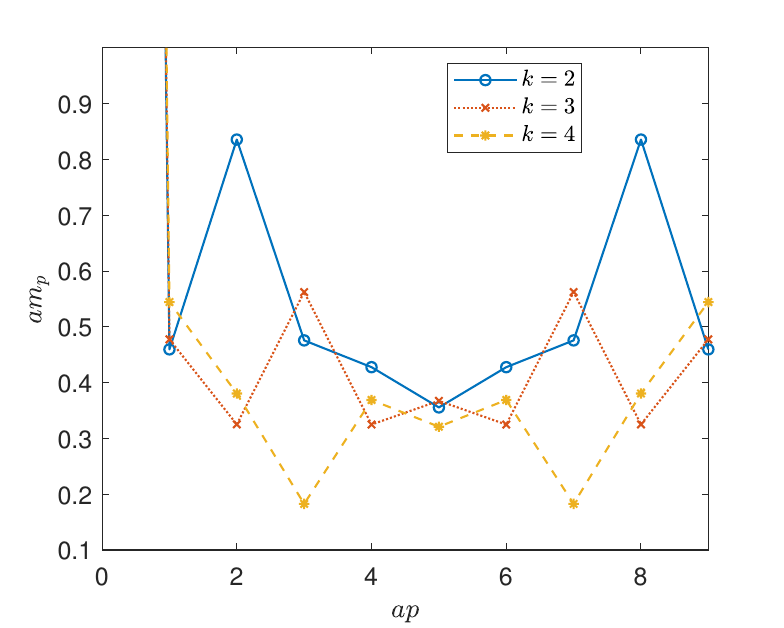}
\caption{\label{fig:mk580uu2}$m_p$ defined in Eq.~(\ref{eq.3.8}) for the $uu$ channels of $\pi$~(left panel) and $a_{1,\perp}$~(right panel) at different $k$~(electric field strength).}
\end{center}
\end{figure}
To verify the oscillatory behavior, $m_p$ for $uu$ channels of $\pi$ and $a_{1,\perp}$ are shown in Fig.~\ref{fig:mk580uu2}.
It can be seen that, these channels follow a similar pattern to the scalar case, with the field-induced oscillations being partially obscured by the additional staggered oscillation, rather than absent.

\subsubsection{\label{sec:sec3.3.3}A summary of the phenomena of the screening mass correlators}

In the study of screening correlators, we find that at low temperatures the neutral $a_0$ screening mass increases with the imaginary electric field, while the $\pi$ channel remains largely insensitive. 
In addition, the charged $a_0$ effective mass exhibits possible spatial oscillations. 
At high temperatures, clear spatial oscillations appear in mesonic effective masses, consistent with similar structures observed in the chiral condensate and the imaginary charge density.

At low temperatures, the system remains confined and no spatial modulation is observed in either the chiral condensate or the charge density, indicating an effectively homogeneous medium. 
In this regime, the oscillatory behavior in the charged~($du$) channel implies that effective-mass oscillations are not solely induced by medium inhomogeneities, but may instead originate from intrinsic phase structures of the quark propagators, particularly due to the mismatch of quark electric charges.

The increase of the $a_0$ screening mass can be further interpreted under the assumption of analyticity near $E_x=0$, where the leading dependence is expected to be even in $E^2$. 
A continuation $E^2 \to -E^2$ would then imply a decrease of the scalar mass in a real electric field, while the pseudo-scalar channel remains approximately unchanged, consistent with its pseudo-Goldstone nature. 
This suggests a possible reduction of the scalar/pseudo-scalar splitting. 
Since real electric fields tend to suppress the chiral condensate and drive the system toward chiral symmetry restoration~\cite{Babansky:1997zh,Klevansky:1989vi,Suganuma:1990nn,Tavares:2019mvq,Cao:2015dya,Ruggieri:2016xww,Ruggieri:2016jrt,Ruggieri:2016lrn}, such a trend is physically plausible. 
However, this argument is heuristic, as real electric fields introduce non-equilibrium effects such as pair production, and the relation between screening and pole masses remains nontrivial.

At high temperatures, quark degrees of freedom respond more directly to the external field. 
The emergence of spatial oscillations in both the chiral condensate and the charge density indicates that the medium itself becomes inhomogeneous. 
Consequently, mesonic correlators probe propagation in a nonuniform background and inherit its spatial structure, leading to oscillatory behavior across all channels.

The response of QCD matter to the external electric field is governed by distinct dynamical regimes: at low temperatures it is dominated by internal hadronic structure, while at high temperatures it reflects medium-level reorganization.

\section{\label{sec:sec4}Summary}

We have studied the response of QCD matter to an external imaginary electric field, focusing on mesonic screening correlators, chiral condensates, charge densities, and the Polyakov loop at low and high temperatures.
The chiral condensate at high temperature, and the Polyakov loop at both temperatures exhibit spatial modulations which are consistent to previous works.

At low temperature, clear plateaus are observed in several channels. 
The scalar screening masses increase with the external field, while the pseudo-scalar masses remain approximately unchanged. 
A weak spatial modulation is observed in the $du$ scalar channel at small field strength. 

At high temperature, the correlators display pronounced oscillatory behavior with frequencies proportional to the quark charges, consistent with the modulation observed in the chiral condensates and charge densities. 
The scalar and pseudo-scalar channels exhibit comparable effective masses, which are generally larger than in the low-temperature case. 
Due to the absence of clear plateaus at high temperature, we restrict the analysis to effective masses rather than fitted screening masses. 
Overall, the results highlight the interplay between external fields, quark charges, and thermal effects in determining the structure of QCD screening correlators.

\appendix

\section{\label{sec:ap1}The mass of meson}

The meson operator is,
\begin{equation}
\begin{split}
&M_{\Gamma,\tau}(h) = \bar{q}_{f_1}(h)\Gamma \otimes \tau^{\dagger} q_{f_2}(h),
\end{split}
\label{eq.ap.1}
\end{equation}
where $h$ is the coordinates of hypercubics, $\tau$ is the matrix in the taste space. And with $q_{a,\alpha}(h)=\sum_{\mu}\sum _{\delta_{\mu}=0,1} \Gamma ^{a\alpha}_{\delta}\chi (h+\delta)/ 8$~\cite{Kluberg-Stern:1983lmr,Morel:1984di}, with,
\begin{equation}
\begin{split}
&\Gamma _{\delta}=\left(\gamma _1\right)^{\delta_x}\left(\gamma _2\right)^{\delta_y}\left(\gamma _3\right)^{\delta_z}\left(\gamma _4\right)^{\delta_{\tau}},
\end{split}
\label{eq.ap.0}
\end{equation}
then Eq.~(\ref{eq.ap.1}) can be rewritten as,
\begin{equation}
\begin{split}
&M_{\Gamma,\tau}(h) = \frac{1}{16}\sum _{n\in h}s_n\bar{\chi}_{f_1}(h+n)\chi_{f_2}(h+n+\delta _n),
\end{split}
\label{eq.ap.2}
\end{equation}
where $s_n=\pm 1$ and $\delta _n$ depend on $\Gamma$ and $\tau$.
The sum over color index is implied in the vector inner product~(note that $\chi$ is a $SU(3)$ vector).
For simplicity, only the cases of `local operators' are considered, where $\Gamma = \tau$ such that one has $\delta _n=0$~(index of $I-IV$ in Table~I of Ref.~\cite{Ishizuka:1993mt}).
In this case, $M_{\Gamma,\tau}$ is denoted as $M_{\Gamma}$ for simplicity.
It can be verified, 
\begin{equation}
\begin{split}
&M_I(h) = \frac{1}{16}\sum _{n\in h} \bar{\chi}_{f_1}(n)\chi_{f_2}(n),\\
&M_{\gamma _{\mu}}(h)= \frac{1}{16}\sum _{n\in h} (-1)^{n_x+n_y+n_z+n_t+n_{\mu}}\bar{\chi}_{f_1}(n)\chi_{f_2}(n),\\
&M_{\gamma _5}(h)= \frac{1}{16}\sum _{n\in h} (-1)^{n_x+n_y+n_z+n_t}\bar{\chi}_{f_1}(n)\chi_{f_2}(n),\\
&M_{\gamma _{\mu}\gamma _5}(h)= \frac{1}{16}\sum _{n\in h} (-1)^{n_{\mu}}\bar{\chi}_{f_1}(n)\chi_{f_2}(n),\\
&M_{\gamma _{\mu}\gamma _{\nu}}(h)= \frac{1}{16}\sum _{n\in h} (-1)^{n_{\mu}+n_{\nu}}\bar{\chi}_{f_1}(n)\chi_{f_2}(n).\\
\end{split}
\label{eq.ap.3}
\end{equation}
Note that Eq.~(\ref{eq.ap.3}) depends on conventions. As a result,
\begin{equation}
\begin{split}
&M_I(h) = \frac{1}{16}\sum _{n\in h} \bar{\chi}_{f_1}(n)\chi_{f_2}(n),\;\;\;\\&M_{\gamma _x\gamma _5}(h)= \frac{1}{16}\sum _{n\in h} (-1)^{n_x}\bar{\chi}_{f_1}(n)\chi_{f_2}(n),\\
&M_{\gamma _x}(h)= \frac{1}{16}\sum _{n\in h} (-1)^{n_y+n_z+n_t}\bar{\chi}_{f_1}(n)\chi_{f_2}(n),\;\;\;\\&M_{\gamma _5}(h)= \frac{1}{16}\sum _{n\in h} (-1)^{n_x+n_y+n_z+n_t}\bar{\chi}_{f_1}(n)\chi_{f_2}(n),\\
&M_{\gamma _j\gamma _5}(h)= \frac{1}{16}\sum _{n\in h} (-1)^{n_j}\bar{\chi}_{f_1}(n)\chi_{f_2}(n),\;\;\;\\&M_{\gamma _x\gamma _j}(h)= \frac{1}{16}\sum _{n\in h} (-1)^{n_x+n_j}\bar{\chi}_{f_1}(n)\chi_{f_2}(n),\\
&M_{\gamma _j\gamma _k}(h)= \frac{1}{16}\sum _{n\in h} (-1)^{n_j+n_k}\bar{\chi}_{f_1}(n)\chi_{f_2}(n),\;\;\;\\&M_{\gamma _l}(h)= \frac{1}{16}\sum _{n\in h} (-1)^{n_x+n_j+n_k}\bar{\chi}_{f_1}(n)\chi_{f_2}(n),\\
\end{split}
\label{eq.ap.4}
\end{equation}
where $j\neq k\neq l \in \{y,z,t\}$.

Taking the mixing of $M_{\gamma _x}(h)$ and $M_{\gamma _5}(h)$ as an example, with translation invariance to combine the two terms~($n_x=m_x=0$ and $n_x=m_x=1$),
\begin{widetext}
\begin{equation}
\begin{split}
&\bar{M}_{\gamma _x}(h)M_{\gamma _x}(0) = \frac{1}{256}\left(\right.\\
&\left.2\sum _{n_{y,z,t}=0,1}W_2(\vec{n})\bar{\chi}_{f_2}(h_x,\vec{h}+\vec{n})\chi_{f_1}(h_x,\vec{h}+\vec{n})\sum _{m_{y,z,t}=0,1}W_2(\vec{m})\bar{\chi}_{f_1}(0,\vec{m})\chi_{f_2}(0,\vec{m})\right.\\
&\left.+\sum _{n_{y,z,t}=0,1}W_2(\vec{n})\bar{\chi}_{f_2}(h_x,\vec{h}+\vec{n})\chi_{f_1}(h_x,\vec{h}+\vec{n})\sum _{m_{y,z,t}=0,1}W_2(\vec{m})\bar{\chi}_{f_1}(1,\vec{m})\chi_{f_2}(1,\vec{m})\right.\\
&\left.+\sum _{n_{y,z,t}=0,1}W_2(\vec{n})\bar{\chi}_{f_2}(h_x+1,\vec{h}+\vec{n})\chi_{f_1}(h_x+1,\vec{h}+\vec{n})\sum _{m_{y,z,t}=0,1}W_2(\vec{m})\bar{\chi}_{f_1}(0,\vec{m})\chi_{f_2}(0,\vec{m})\right),\\
&\bar{M}_{\gamma _5}(h)M_{\gamma _5}(0) = \frac{1}{256}\left(\right.\\
&\left.2\sum _{n_{y,z,t}=0,1}W_2(\vec{n})\bar{\chi}_{f_2}(h_x,\vec{h}+\vec{n})\chi_{f_1}(h_x,\vec{h}+\vec{n})\sum _{m_{y,z,t}=0,1}W_2(\vec{m})\bar{\chi}_{f_1}(0,\vec{m})\chi_{f_2}(0,\vec{m})\right.\\
&\left.-\sum _{n_{y,z,t}=0,1}W_2(\vec{n})\bar{\chi}_{f_2}(h_x,\vec{h}+\vec{n})\chi_{f_1}(h_x,\vec{h}+\vec{n})\sum _{m_{y,z,t}=0,1}W_2(\vec{m})\bar{\chi}_{f_1}(1,\vec{m})\chi_{f_2}(1,\vec{m})\right.\\
&\left.-\sum _{n_{y,z,t}=0,1}W_2(\vec{n})\bar{\chi}_{f_2}(h_x+1,\vec{h}+\vec{n})\chi_{f_1}(h_x+1,\vec{h}+\vec{n})\sum _{m_{y,z,t}=0,1}W_2(\vec{m})\bar{\chi}_{f_1}(0,\vec{m})\chi_{f_2}(0,\vec{m})\right),\\
\end{split}
\label{eq.ap.5}
\end{equation}
\end{widetext}
where $W_2(\vec{n})=(-1)^{n_y+n_z+n_t}$, and $\vec{h}$, $\vec{m}$, and $\vec{n}$ are defined in $y-z-t$ space.
In Eq.~(\ref{eq.ap.5}), we denote $\chi (n)=\chi (n_x, \vec{n})$.

Still use translation invariance and use $W(\vec{n})W(\vec{m})=W(\vec{n}+\vec{m})=W(\vec{n}-\vec{m})$,
\begin{equation}
\begin{split}
&\sum _{n_{y,z,t}=0,1}W_2(\vec{n})\bar{\chi}_{f_2}(h_x,\vec{h}+\vec{n})\chi_{f_1}(h_x,\vec{h}+\vec{n})\\
&\times \sum _{m_{y,z,t}=0,1}W_S(\vec{m})\bar{\chi}_{f_1}(0,\vec{m})\chi_{f_2}(0,\vec{m})\\
&=\sum _{n_{y,z,t}=0,1}\sum _{m_{y,z,t}=0,1}W_2(\vec{n}-\vec{m})\times \\
&\bar{\chi}_{f_2}(h_x,\vec{h}+\vec{n}-\vec{m})\chi_{f_1}(h_x,\vec{h}+\vec{n}-\vec{m})\bar{\chi}_{f_1}(0,0)\chi_{f_2}(0,0)
\end{split}
\label{eq.ap.6}
\end{equation}
For each $\vec{h}$, there are $64$ terms where $8$ of which $|\vec{n}-\vec{m}|=0$, $24$ of which $|\vec{n}-\vec{m}|=1$, $24$ of which $|\vec{n}-\vec{m}|=\sqrt{2}$, $8$ of which $|\vec{n}-\vec{m}|=\sqrt{3}$.
One can reassign these terms, retaining only those where the $\vec{n}-\vec{m}$ component is non-negative, assigning those with a negative $\vec{n}-\vec{m}$ component to the adjacent hypercubic, while simultaneously receiving from the adjacent hypercubic those terms where the $\vec{n}_n-\vec{m}_n$ component is negative but $\vec{n}-\vec{m}$ is negative (where $\vec{n}_n$ and $\vec{m}_n$ are the vectors with the coordinate of the adjacent hypercubic as the origin of the frame, $\vec{n}$ and $\vec{m}$ are the vectors with the coordinate of the this hypercubic as the origin of the frame).
As a result, there are eight terms where one of which $|\vec{n}-\vec{m}|=0$, three of which $|\vec{n}-\vec{m}|=1$, three of which $|\vec{n}-\vec{m}|=\sqrt{2}$, one of which $|\vec{n}-\vec{m}|=\sqrt{3}$, so by redefinition,
\begin{equation}
\begin{split}
&\sum _{n_{y,z,t}=0,1}W_2(\vec{n})\bar{\chi}_{f_2}(h_x, \vec{h}+\vec{n})\chi_{f_1}(h_x, \vec{h}+\vec{n})\\
&\times \sum _{m_{y,z,t}=0,1}W_2(\vec{m})\bar{\chi}_{f_1}(0, \vec{m})\chi_{f_2}(0, \vec{m})\\
&\to 8\sum _{n_{y,z,t}=0,1}W_2(\vec{n})\\
&\times \bar{\chi}_{f_2}(h_x, \vec{h}+\vec{n})\chi_{f_1}(h_x, \vec{h}+\vec{n})\bar{\chi}_{f_1}(0,0)\chi_{f_2}(0,0).
\end{split}
\label{eq.ap.7}
\end{equation}

Defining $C_{\Gamma}(h)=\bar{M}_{\Gamma}(h)M_{\Gamma}(0)$,
\begin{equation}
\begin{split}
&\left(C_{\gamma _x}(h)+C_{\gamma _5}(h)\right)=\frac{1}{8}\sum _{n_{y,z,t}=0,1}W_2(\vec{n})\\
&\times \bar{\chi}_{f_2}(h_x,\vec{h}+\vec{n})\chi_{f_1}(h_x,\vec{h}+\vec{n})\bar{\chi}_{f_1}(0,0)\chi_{f_2}(0,0)\\
&\left(C_{\gamma _x}(h)-C_{\gamma _5}(h)\right)=\frac{1}{16}\sum _{n_{y,z,t}=0,1}W_2(\vec{n})\\
&\times \bar{\chi}_{f_2}(h_x+1,\vec{h}+\vec{n})\chi_{f_1}(h_x+1,\vec{h}+\vec{n})\bar{\chi}_{f_1}(0,0)\chi_{f_2}(0,0)\\
&+\frac{1}{16}\sum _{n_{y,z,t}=0,1}W_2(\vec{n})\\
&\times \bar{\chi}_{f_2}(h_x-1,\vec{h}+\vec{n})\chi_{f_1}(h_x-1,\vec{h}+\vec{n})\bar{\chi}_{f_1}(0,0)\chi_{f_2}(0,0)\\
\end{split}
\label{eq.ap.8}
\end{equation}
Since $h$ are all even sites and $W_2(\vec{n})=W_2(\vec{h}+\vec{n})$, similarly as above, if we put the $h_x-1$ to the definition of $C_{\Gamma}(h_x-2, \vec{h})$, 
\begin{equation}
\begin{split}
&\left(C_{\gamma _x}(h)+C_{\gamma _5}(h)\right)=\frac{1}{8}\sum _{\vec{n}}W_2(\vec{n})\\
&\times \bar{\chi}_{f_2}(h_x,\vec{n})\chi_{f_1}(h_x,\vec{h}+\vec{n})\bar{\chi}_{f_1}(0,0)\chi_{f_2}(0,0)\\
&\left(C_{\gamma _x}(h)-C_{\gamma _5}(h)\right)=\frac{1}{8}\sum _{\vec{n}}W_2(\vec{n})\\
&\times \bar{\chi}_{f_2}(h_x+1,\vec{n})\chi_{f_1}(h_x+1,\vec{h}+\vec{n})\bar{\chi}_{f_1}(0,0)\chi_{f_2}(0,0)\\
\end{split}
\label{eq.ap.9}
\end{equation}
Defining,
\begin{equation}
\begin{split}
&C^{(i)}_{f_1f_2}(n_x, \vec{n})=W_i\bar{\chi}_{f_2}(n_x, \vec{n})\chi_{f_1}(h_x, \vec{n})\bar{\chi}_{f_1}(0,0)\chi_{f_2}(0,0),
\end{split}
\label{eq.ap.10}
\end{equation}
it can be seen that, with,
\begin{equation}
\begin{split}
&W_1=1,\;\;W_2=(-1)^{n_y+n_z+n_t},\;\;\\&W_3=(-1)^{n_j},\;\;W_4=(-1)^{n_j},\\
\end{split}
\label{eq.ap.11}
\end{equation}
one has,
\begin{equation}
\begin{split}
&C^{(1)}_{f_1f_2}(n_x,\vec{n})=C_I(h) + (-1)^{n_x} C_{\gamma_x\gamma _5}(h),\\
&C^{(2)}_{f_1f_2}(n_x,\vec{n})=C_{\gamma _x}(h) + (-1)^{n_x} C_{\gamma _5}(h),\\
&C^{(3)}_{f_1f_2}(n_x,\vec{n})=C_{\gamma _j\gamma _5}(h) + (-1)^{n_x} C_{\gamma_x\gamma _j}(h),\\
&C^{(4)}_{f_1f_2}(n_x,\vec{n})=C_{\gamma _j\gamma _k}(h) + (-1)^{n_t} C_{\gamma_l}(h),\\
\end{split}
\label{eq.ap.12}
\end{equation}
which are the four channels in Ref.~\cite{Gottlieb:1988gr} but in the case of screening mass.

With the Gaussian integration of Grassman numbers, 
\begin{equation}
\begin{split}
&\langle \bar{\chi} _{i_1}\chi _{j_1} \bar{\chi} _{i_2} \chi _{j_2}  \rangle = D^{-1}_{j_1i_1} D^{-1}_{j_2i_2} - D^{-1}_{j_1i_2} D^{-1}_{j_2i_1}.\\
\end{split}
\label{eq.ap.13}
\end{equation}
In this work, we restrict to the connected part~(the second term in the right hand side) of the correlators, neglecting disconnected contributions~(the first term in the right hand side), which are computationally expensive and vanish in flavor non-singlet channels, one has,
\begin{equation}
\begin{split}
&\sum _{c_1,c_2}\langle \bar{\chi}^{c_1}_{f_1}(n)\chi^{c_1}_{f_2}(n)\bar{\chi}^{c_2}_{f_1}(0,0)\chi^{c_2}_{f_2}(0,0) \rangle \\
&= - \sum _{c_1,c_2} D^{-1}_{f_1}(n,c_1|0,c_2) D^{-1}_{f_2}(0,c_2|n,c_1)\\
&= - \sum _{c_1,c_2} D^{-1}_{f_1}(n,c_1|0,c_2) \left((D^{\dagger})^{-1}_{f_2}\right)^*(n,c_1|0,c_2),\\
\end{split}
\label{eq.ap.14}
\end{equation}

The calculation of $D^{-1}_{f_1}(n,c_1|0,c_2)$ is,
\begin{equation}
\begin{split}
&\langle v_{n,c_1}|D^{-1}|v_{0,c_2}\rangle\\
\end{split}
\label{eq.ap.15}
\end{equation}
where $v_i$ is a vector with only the $i$-th component been non-zero and $v_i=1$. Therefore,
\begin{equation}
\begin{split}
&D^{-1}(n,c_1|0,c_2) = \langle v_{n,c_1}|D^{\dagger}(DD^{\dagger})^{-1}|v_{0,c_2}\rangle,\\
&(D^{\dagger})^{-1}(n,c_1|0,c_2) = \langle v_{n,c_1}|D(DD^{\dagger})^{-1}|v_{0,c_2}\rangle.\\
\end{split}
\label{eq.ap.16}
\end{equation}

The $v_{n,c_1}$ and $v_{0,c_2}$ are often called as the sink and source, respectively.
In this work, we use a point source, but sum over the results of a whole x-slice $v_{n,c_1}$ to calculate $C^{(i)}_{f_1f_2}(n_x)=\sum _{\vec{n}}C^{(i)}_{f_1f_2}(n_x, \vec{n})$.

After $C^{(i)}_{f_1f_2}(n_x)$ is calculated, it can be fitted according to,
\begin{equation}
\begin{split}
&C^{(i)}_{f_1f_2}(n_x)=A_+\left(e^{-m_+n_x}+e^{-m_+(L_x-n_x)}\right)\\
&+(-1)^{n_x}A_-\left(e^{-m_-n_x}+e^{-m_-(L_x-n_x)}\right).\\
\end{split}
\label{eq.ap.17}
\end{equation}

\begin{table*}
\begin{center}
\begin{tabular}{c|c|cc|cc}
\hline    
$i$ & $W_i(\vec{n})$ & Part. interp. & $\Gamma$ & Opp. parity & $\Gamma$ \\
\hline
$1$ & $1$ & scalar ($a_0$) & $1$ & axial vector ($a_{1,\parallel}$) & $\gamma _x \gamma _5$ \\
$2$ & $(-1)^{n_y+n_z+n_t}$ & vector ($\rho_{\parallel}$) & $\gamma _x$ & pseudo-scalar ($\pi$) & $\gamma _5$ \\
$3$ & $(-1)^{n_y}+(-1)^{n_z}$ & axial vector ($a_{1,\perp}$) & $\gamma _{y,z}\gamma _5$ & tensor & $\gamma _x\gamma _{y,z}$ \\
$4$ & $(-1)^{n_t}$ & temporal axial vector & $\gamma _4\gamma _5$ & tensor & $\gamma _x\gamma _4$ \\
$5$ & $(-1)^{n_y+n_z}$ & tensor & $\gamma _y\gamma _z$ & temporal vector ($\rho _4$) & $\gamma _4$ \\
$6$ & $(-1)^{n_y+n_t}+(-1)^{n_z+n_t}$ & tensor & $\gamma _{y,z}\gamma _4$ & vector ($\rho_{\perp}$) & $\gamma _{y,z}$ \\
\hline
\end{tabular}
\end{center}
\caption{\label{tab.mesonchannels}Corresponds of different sign functions and different meson channels.
The labels `parallel' and `perpendicular' refer to orientations relative to the screening direction $x$, and do not imply full rotational symmetry.
For local operators, there are four channels, the six channels are listed due to the splitting of `parallel' and `perpendicular' cases.
Note that the $W_i(n)$ functions depend on conventions.
We use a convention same as Refs.~\cite{Ishizuka:1993mt,Golterman:1985dz,Altmeyer:1992dd}, and has a $(-1)^{n_x+n_y+n_z+n_{\tau}}$ difference from Refs.~\cite{Gottlieb:1988gr,Cheng:2010fe}.}
\end{table*}
The different channels are summarized in Table~\ref{tab.mesonchannels}

\begin{acknowledgments}
This work was supported in part by the National Natural Science Foundation of China under Grants Nos. 11875157 and 12147214, and the Natural Science Foundation of the Liaoning Scientific Committee No.~LJKMZ20221431.
\end{acknowledgments}

\bibliography{MesonMass}

\end{document}